# Regulating Gatekeeper AI and Data: Transparency, Access, and Fairness under the DMA, the GDPR, and beyond

Authors: Philipp Hacker[*], Johann Cordes[†] and Janina Rochon[‡]

**Abstract**:

Artificial intelligence (AI) is not only increasingly used in business and administration contexts, but a race for its regulation is also underway, with the EU spearheading the efforts. Contrary to existing literature, this article suggests that the most far-reaching and effective EU rules for AI applications in the digital economy will not be contained in the proposed AI Act–but in the Digital Markets Act (DMA).

We analyze the impact of the DMA and related EU acts on AI models and underlying data across four key areas: disclosure requirements; the regulation of AI training data; access rules; and the regime for fair rankings. We demonstrate that fairness, under the DMA, goes beyond traditionally protected categories of non-discrimination law on which scholarship at the intersection of AI and law has focused on. Rather, we draw on competition law and the FRAND criteria known from intellectual property law to interpret and refine the DMA provisions on fair rankings. Moreover, we show how, based on CJEU jurisprudence, a coherent interpretation of the concept of non-discrimination in both traditional non-discrimination and competition law may be found. The final part sketches proposals for a comprehensive framework of transparency, access, and fairness under the DMA and beyond.

---

[*] Professor Dr. Philipp Hacker, LL.M. (Yale), Chair for Law and Ethics of the Digital Society, European New School of Digital Studies, European University Viadrina. We are grateful for research assistance by Sarah Großheim and Marco Mauer.
[†] Research Fellow, Chair for Law and Ethics of the Digital Society, European New School of Digital Studies, European University Viadrina.
[‡] Research Fellow, Chair for Law and Ethics of the Digital Society, European New School of Digital Studies, European University Viadrina.

## Table of Contents

## I. Introduction

Artificial intelligence (AI)[1] and, in particular, machine learning (ML)[2] are not only at the center of countless economic applications,[3] but have also triggered a broad regulatory debate because of their partial lack of transparency[4] and tendencies to perpetuate discrimination[5] in some models and scenarios.[6] At the EU level, the supposedly main pillar of AI regulation was proposed by the EU Commission in April 2021:[7] the AI Act.[8] It is currently hotly debated both in the EU Council and Parliament[9] and will probably be adopted in late 2023 or 2024.[10] In addition, with the recently enacted Digital Services Act (DSA) and Digital Markets Act (DMA), the EU is pursuing the aim of regulating larger platforms to ensure a more competitive environment for smaller providers and to support the creation of a safer digital space.[11] In September 2022, the final, missing piece of European AI regulation was unveiled with the Commission proposal of two directives[12] concerning AI and software liability.[13]

---

[1] In this paper, AI is understood based on the definition in Article 3(1) and Recitals 6a and 6b AI Act. This definition is not far from perfect, see Stuart J. Russell and Peter Norvig, Artificial Intelligence: A Modern Approach (3rd Global ed. edn, Pearson Education, Inc. 2016) 5; David L. Poole and Alan K. Mackworth, Artificial Intelligence: Foundations of Computational Agents (2nd ed. edn, Cambridge University Press 2017), 3-7; Matt O'Shaughnessy, 'One of the Biggest Problems in Regulating AI Is Agreeing on a Definition' Carnegie Endowment <https://carnegieendowment.org/2022/10/06/one-of-biggest-problems-in-regulating-ai-is-agreeing-on-definition-pub-88100> accessed 6 December 2022; Philipp Hacker, 'The European AI Liability Directives--Critique of a Half-Hearted Approach and Lessons for the Future' (2022) arXiv preprint arXiv:221113960, 11-12. However, it constitutes a workable definition for the purposes of this paper.

[2] For a definition of ML, see Tom M. Mitchell, Machine Learning (1st ed., 1997) 2: "A computer program is said to learn from experience E with respect to some class of tasks T and performance measure P, if its performance at tasks in T, as measured by P, improves with experience E."

[3] Ian Goodfellow, Yoshua Bengio and Aaron Courville, Deep Learning (2016) 23 et seq.; Javaid et al., 'Artificial Intelligence Applications for Industry 4.0: A Literature-Based Study' (2022) 7 Journal of Industrial Integration and Management 83.

[4] Zachary C Lipton, 'The mythos of model interpretability: In machine learning, the concept of interpretability is both important and slippery' (2018) 16 Queue 31.

[5] Ferrer et al., 'Bias and Discrimination in AI: a cross-disciplinary perspective' (2021) 40 IEEE Technology and Society Magazine 72.

[6] See, e.g., Chris Reed, 'How should we regulate artificial intelligence?' (2018) 376.2128 Philosophical Transactions of the Royal Society A: Mathematical, Physical and Engineering Sciences, Article 20170360; Miriam Buiten, 'Towards intelligent regulation of artificial intelligence' (2019) 10 European Journal of Risk Regulation 41; Herbert Zech, 'Entscheidungen digitaler autonomer Systeme: Empfehlen sich Regelungen zu Verantwortung und Haftung?' (Deutscher Juristentag, Bonn, 2020); Philipp Hacker, 'Europäische und nationale Regulierung von Künstlicher Intelligenz' (2020) Neue Juristische Wochenzeitschrift 2142; Nathalie Smuha, 'From a 'race to AI' to a 'race to AI regulation': regulatory competition for artificial intelligence' (2021) 13 Law, Innovation and Technology 57; Martin Ebers M and others, 'The European commission's proposal for an artificial intelligence act—a critical assessment by members of the robotics and AI law society (RAILS)' (2021) 4 J 589; Christiane Wendehorst and Jakob Hirtenlehner, 'Outlook on the Future Regulatory Requirements for AI in Europe' (2022) Working Paper 2022, https://ssrncom/abstract=4093016.

[7] For analysis and critique, see, e.g., Michael Veale and Frederik Zuiderveen Borgesius, 'Demystifying the Draft EU Artificial Intelligence Act—Analysing the good, the bad, and the unclear elements of the proposed approach' (2021) 22 Computer Law Review International 97, as well as some of the references in n. 6.

[8] European Commission, Proposal for a Regulation of the European Parliament and of the Council laying down harmonised rules on Artificial Intelligence, COM(2021) 206 final; all references to the AI Act are to the following document: Council of the EU, Interinstitutional File: 2021/0106(COD), General Approach (= final version of the Council compromise text) of Nov. 25, 2022, Doc. No. 14954/22.

[9] Luca Bertuzzi, 'Artificial Intelligence definition, governance on MEPs' menu' (EURACTIV 10 November 2022) <https://www.euractiv.com/section/digital/news/artificial-intelligence-definition-governance-on-meps-menu> accessed 10 November 2022.

[10] See, e.g., the remarks by MEP Axel Voss in Luca Bertuzzi, The new liability rules for AI, 108 The Tech Brief, Euractiv (EURACTIVE 30 September 2022) <https://www.euractiv.com/section/digital/podcast/the-new-liability-rules-for-ai/>, accessed 9 November 2022.

With this package, the EU continues to build and refine the regulatory framework for the Digital Single Market. While much of the legal and technical debate concerning AI regulation has focused on the AI Act and the related AI liability provisions,[14] this paper seeks to show that the most far-reaching, most overlooked, but potentially also most effective regulatory constraints for AI are ultimately erected by the DMA. The act contains numerous regulations for so-called gatekeepers, large online platforms such as Google or Amazon, which exceed certain quantitative benchmarks and are thus essential for access to digital markets (Article 3 DMA). The possibility of fines – unmatched so far in any area of EU business law–underscore the importance of the DMA for legal and digital practice: authorities may punish a violation of the DMA with a fine up to 10% of the total annual global turnover (Article 30(1) DMA).

However, the DMA is currently being discussed primarily as an instrument to contain the market power of large online platforms.[15] In contrast, scholars have paid considerably less attention to the fact that the DMA contains rules that will probably have greater significance for the application of machine learning in the EU than the regulations provided for in the AI Act.[16] The central measures of the latter apply only to high-risk applications, which are, however, rare when ML deployments in the digital economy are considered.[17] The ML applications of Google, Amazon and other large platforms will *not* be subject to the strict rules of the AI Act–but to those of the DMA. AI and ML power the core of the business and competitive edge of all the large platforms: the ranking and scoring models used to present customers optimized, and often personalized, lists of items in return for a specific query. The quality of these rankings attracts customers, who in turn attract advertisers. Rankings are, therefore, key to platform success.

---

[11] Regulation (EU) 2022/2065 of the European Parliament and of the Council of 19 October 2022 on a Single Market for Digital Services, OJ L277/1 (DSA); Regulation (EU) 2022/1925 of the European Parliament and of the Council of 14 September 2022 on contestable and fair markets in the digital sector, OJ L265/1 (DMA).

[12] European Commission, Proposal for a Directive of the European Parliament and of the Council on Liability for Defective Products, COM(2022) 495 final [PLD Proposal]; European Commission, Proposal for a Directive of the European Parliament and of the Council on adapting non-contractual civil liability rules to artificial intelligence, COM(2022) 496 final [AILD Proposal].

[13] See, e.g., Philipp Hacker, 'The European AI Liability Directives--Critique of a Half-Hearted Approach and Lessons for the Future' (2022) arXiv preprint arXiv:221113960; Gerald Spindler, 'Die Vorschläge der EU-Kommission zu einer neuen Produkthaftung und zur Haftung von Herstellern und Betreibern Künstlicher Intelligenz' (2022) Computer und Recht 689.

[14] See, e.g., Michael Veale and Frederik Zuiderveen Borgesius, 'Demystifying the Draft EU Artificial Intelligence Act—Analysing the good, the bad, and the unclear elements of the proposed approach' (2021) 22 Computer Law Review International 97; Natali Helberger and Nicholas Diakopoulos, 'The European AI Act and how it matters for research into AI in media and journalism' (2022) Digital Journalism 1; Francesco Sovrano et al., 'Metrics, Explainability and the European AI Act Proposal' (2022) 5 J 126; Natali Smuha et al., 'How the EU can achieve legally trustworthy AI: a response to the European commission's proposal for an artificial intelligence act' (2021), https://papers.ssrn.com/sol3/papers.cfm?abstract_id=3899991; Philipp Hacker and Jan-Hendrik Passoth, 'Varieties of AI Explanations Under the Law. From the GDPR to the AIA, and Beyond' in Holzinger et al. (eds), xxAI – Beyond Explainable AI, International Workshop on Extending Explainable AI Beyond Deep Models and Classifiers (Springer 2022), 343; Margot E Kaminski, 'Regulating the Risks of AI' (2023) 103 Boston University Law Review (forthcoming); Philipp Hacker, 'A legal framework for AI training data—from first principles to the Artificial Intelligence Act' (2021) 13 Law, Innovation and Technology 257.

[15] See only Eifert et al., 'Taming the giants: The DMA/DSA package' (2021) 58 Common Market Law Review 987; Johann Laux, Sandra Wachter, Brent Mittelstadt, 'Taming the Few: Platform Regulation, Independent Audits, and the Risks of Capture Created by the DMA and DSA' (2021) 43 Computer Law & Security Review 105613; Rupprecht Podszun, Philipp Bongartz and Sarah Langenstein, 'The Digital Markets Act: Moving from Competition Law to Regulation for Large Gatekeepers' (2021) 11 EuCML 60; Antonio Davola and Gianclaudio Malgieri, 'Data, Power and Competition Law: The (Im) Possible Mission of the DMA?' (2022) 2023 Research in Law and Economics, Forthcoming, https://papers.ssrn.com/sol3/papers.cfm?abstract_id=4242574.

[16] See, e.g., Meike Zehlike et al., 'Beyond Incompatibility: Interpolation between Mutually Exclusive Fairness Criteria in Classification Problems' (2022) arXiv:2212.00469; Philipp Hacker, 'KI und DMA' (2022) 75 Gewerblicher Rechtsschutz und Urheberrecht 1278.

For example, when you search for a pair of shoes on Amazon, the resulting offers (the ranking) is created by a powerful ML learning-to-rank model that takes a variety of factors (so-called 'features') into account.[18] Similarly, Google answers queries based on complex AI models.[19] The competitive edge of large platforms lies precisely in their capability to build such rankings in a fast, meaningful, and personalized way.

For the first time in history, the EU has now enacted, with the DMA, provisions explicitly and specifically regulating such rankings and other AI applications of gatekeepers. Due to the centrality of these applications to gatekeeper technology implementation and business, the provisions are likely to have a profound effect on the core systems of the digital economy. Although the term 'artificial intelligence' is absent from the entire DMA, the article identifies four regulatory complexes in the DMA that will significantly change the legal framework for AI in the EU: prerequisites for the creation of fair rankings; information requirements; regulation of training data; and access rights.

In doing so, the article proceeds in several steps. First, we focus on the provisions that have the potentially most far-reaching effect on the use of AI applications by gatekeepers: those governing the fairness, i.e., the very nature and order of rankings (II.). This section also includes an overview of the existing regulations and practices in the field, as well as an evaluation of the newly proposed rules. Next, the paper examines and critiques the new DMA rules regulating training data (III.), access rights (IV.), and information requirements (V.). Concerning training data and transparency, the DMA rules are contrasted with the relevant counterparts in the AI Act. Regarding access rights, we compare the DMA provisions with the current proposal of an EU Data Act. The final substantive part maps out a framework for fairness, transparency and access in gatekeeper AI going forward (VI.). Section VII. concludes.

## II. Regulating AI-based rankings

Rankings have infiltrated virtually all areas of our lives. The prominence given to goods or services offered by online providers undeniably has the power to steer and potentially control our choices.[20] The vast amount of offerings fighting for our attention has created the need for an intermediary, who supports us in our decision-making process by organizing and prioritizing them. This directly results from the digital economy we live in. By controlling the demand, search and buying behavior of individuals, rankings make up the boiler room of this new economy. Unsurprisingly, therefore, the business models of many tech-giants use them as the basis for their competitive advantage.[21] This competitive advantage is even more prominent where the entity also directly sells or provides certain services or products, in addition to their role as an intermediary (vertical integration).[22]

---

[17] They are limited to medical AI, credit scoring, life and health insurance, and employment; see Annexes II and III of the AI Act Proposal.
[18] See for the "A9" algorithm V. Sandeep, B. Pohutezhini, 'The e-commerce revolution of amazon. com' (2019) 6 Splint Internationals Journal of Professionals 33, 37; Maio/Re, 2 IJTB 8, 10 (2020); Trutz Fries, 'Amazon A9 - Amazon's ranking algorithm explained' (Amalytix 10 December 2020) <https://www.amalytix.com/en/knowledge/seo/amazon-alogrithm-a9/>, accessed 4 December 2022.
[19] Ao-Jan Su et al., 'How to improve your Google ranking: Myths and reality. 2010 IEEE/WIC/ACM International Conference on Web Intelligence and Intelligent Agent Technology' (2010), 50; Matt G. Southern, 'Is RankBrain A Ranking Factor In Google Search? ' (SEJ, 5 October 2022) <https://www.searchenginejournal.com/ranking-factors/rankbrain-ranking-factor> accessed 4 December 2022.

The European Commission has identified the immense impact the control of rankings has on market power as a potential problem and, therefore, included specific regulations on their creation and implementation in the DMA. After a short introduction to the differences between AI-based and traditional rankings (1.), this section will provide an overview of the existing legal framework applicable to rankings and contrast it in detail with the new rules introduced by the DMA. In doing so, the section covers the most important regulatory dimensions of rankings: transparency (2.), accuracy and rectification (3.), and fairness/non-discrimination (4.).

### 1. Rankings: a cornerstone of the digital economy

A ranking is an ordered list[23] typically displayed in response to a search entry.[24] In the DMA, ranking is understood as the "relative prominence given to goods or services" or "the relevance given to search results by online search engines [...]" (cf. Article 2(22) DMA), irrespective of the technical means used for such presentation.[25] Such a ranking is created, for example, when a consumer searches for a refrigerator on a comparison or online shopping platform. In this sense, rankings establish a pre-selection of goods or services to facilitate consumers' purchasing decisions.[26]

#### a) Traditional versus AI-based rankings

In theory, a ranking can be manually created: a person could put specific entries on a list in a particular order based on specific predetermined criteria. One may conceive a shopkeeper who sorts certain goods, e.g., hard drives, according to their writing speed, then notes this on a piece of paper and displays it visibly in her shop to help her customers decide. Indeed, according to Article 2(22) DMA, relative prominence or relevance qualifies a list as a ranking, regardless of the technical means by which one creates it. The legal definition of the DMA says nothing about the technical means themselves. To automate this process, one could also record the writing speed of the hard drives in a machine-readable way and rank it using a simple sorting algorithm.[27]

---

[20] See, e.g., Tat-How Teh, Julian Wright, 'Intermediation and steering: Competition in prices and commissions' (2022) 14 AEJ: Microeconomics 281 (2022).
[21] See n. 18 and 19.
[22] See, e.g., Jorge Padilla, Joe Perkins, Salvatore Piccolo, 'Self-Preferencing in Markets with Vertically Integrated Gatekeeper Platforms' (2022) 70 The Journal of Industrial Economics 371; Andrei Hagiu, Tat-How Teh and Julian Wright, 'Should platforms be allowed to sell on their own marketplaces?' (2022) 53 The RAND Journal of Economics 297.
[23] Racula Ursu, 'The Power of Rankings: Quantifying the Effect of Rankings on Online Consumer Search and Purchase Decisions' (2018) 37 Marketing Science 530; Mohri/Rostamizadeh/Talwalkar, Foundations of Machine Learning, (2nd edition, The MIT Press 2018) 3.
[24] Philipp Hacker, 'KI und DMA' (2022) 75 Gewerblicher Rechtsschutz und Urheberrecht 1278, 1281.
[25] The definition essentially corresponds to that of Article 2(8) of Regulation (EU) 2019/1150 on fairness and transparency for business users of online intermediary services (P2B Regulation); on the P2B Regulation see below II.2.a).
[26] Racula Ursu, 'The Power of Rankings: Quantifying the Effect of Rankings on Online Consumer Search and Purchase Decisions' (2018) 37 Marketing Science 530.

With vast amounts of data and accompanying large amounts of potentially relevant criteria for ordering, it has become possible and popular to resort to machine learning techniques to order the data and items efficiently.[28] Such rankings thus differ from the traditional rankings described above primarily in how the data is processed and rankings are personalized. With the release of ChatGPT[29] and implementing this model in Microsoft's search engine Bing,[30] there are now other ways to get a ranking: via a dialog with a chatbot.

### b) The centrality of rankings for online platforms

The data processing creates specific legal challenges of data protection, transparency, accuracy, and fairness, which the law must address. Regulating for transparent, accurate, and fair rankings is crucial for three reasons. First, as mentioned, for many gatekeepers, the ability to produce high-quality rankings often constitutes a pivotal point of their business model, as their competitive advantage often lies precisely in producing fast and reliable, relevant rankings.[31] This is also illustrated by the example of Microsoft's Bing search engine. By implementing ChatGPT,[32] the search engine has been seriously competing with its more popular competitor Alphabet and its search engine Google again for quite some time. It is not all that new for search engines to leverage the capabilities of language models; Google, for example, has been using language models to better understand user queries for quite some time.[33] However, with implementing large generative AI models in search engines, it is now also possible to have a product or service displayed via a chat interface. Second, as empirical studies show, the ranking order has a considerable impact on the decision made by the consumer.[34] Finally, gatekeepers typically use machine learning techniques for the generation of such rankings.[35] Nevertheless, many applications of rankings in the digital economy, particularly in e-commerce or social media, do not qualify as high risk under the AI Act,[36] and will therefore only be regulated by existing EU legislation, including the new Article 6(5) DMA. In this respect, rankings generated by large generative AI models are an exception.[37] However, large generative AI models do not create all rankings.

---

[27] See, e.g., AD Mishra and D Garg, 'Selection of best sorting algorithm' (2008) 2 International Journal of Intelligent Information Processing 363; however, the sorting algorithms, that can be used differ, sometimes considerably, in their efficiency, i.e., for example in terms of their runtime and the memory they require; for an overview see, e.g., Sedgewick/Wayne, Algorithms (4th edition, Addison Wesley 2011) Chapter 2.
[28] Google, for example, uses ML techniques to understand search queries best and provide the most relevant answers, see Barry Schwartz, 'How Google uses artificial intelligence In Google Search. From RankBrain, Neural Matching, BERT and MUM - here is how Google uses AI for understanding language for query, content and ranking purposes' (Search Engine Land, 3 February 2022) <https://searchengineland.com/how-google-uses-artificial-intelligence-in-google-search-379746> accessed 6 December 2022; see also n. 19; Amazon uses such techniques to, among other things, suggest ads based on previous search behavior of potential buyers, see n. 18.
[29] OpenAI, 'Introducing ChatGPT' (Official OpenAI Blog, 30 November 2022) <https://openai.com/blog/chatgpt> accessed 5 May 2023.
[30] Yusuf Mehdi, 'Reinventing search with a new AI-powered Microsoft Bing and Edge, your copilot for the web' (Official Microsoft Blog, 7 February 2023) <https://blogs.microsoft.com/blog/2023/02/07/reinventing-search-with-a-new-ai-powered-microsoft-bing-and-edge-your-copilot-for-the-web/> accessed 27 March 2023.
[31] Martens, 'An Economic Policy Perspective on Online Platforms' (2016) Institute for Prospective Technological Studies Digital Economy Working Paper 2016/05, 4, 20 et seqq.
[32] See n 30.
[33] See n 28 and also n. 19.

It is important, however, to point out that the DMA rules for rankings, such as Article 6(5) DMA, are based on and closely related to previously existing provisions. An increasing number of regulations are dedicated to the transparency (II.2.), accuracy (II.3.), and fairness (II.4.) of online rankings.

## 2. Transparency of rankings

A stumbling block toward effective functioning and regulation of rankings is the information asymmetry between the platforms and all other stakeholders.[38] In recent years, the EU has therefore adopted several critical regulations on the transparency of rankings. These include the P2B Regulation,[39] the Consumer Rights Directive (CRD)[40] and the Unfair Commercial Practices Directive (UCPD),[41] as updated by the Omnibus Directive,[42] respectively; the GDPR;[43] and the DMA, as well as the proposed AI Act.

### a) P2B Regulation

The P2B Regulation targets online brokerage services and search engines and is not limited to market-dominant undertakings. According to Article 5(1) and (2) of the P2B Regulation, both entities must disclose the main parameters that determine the ranking.[44] Mediation services

---

[34] Racula Ursu, 'The Power of Rankings: Quantifying the Effect of Rankings on Online Consumer Search and Purchase Decisions' (2018) 37 Marketing Science 530; Derakhshan et al., 'Product Ranking on Online Platforms' (2022) 68 Management Science 4024, 4028.
[35] Tie-Yan Liu, 'Learning to Rank for Information Retrieval' (2009) 3 Foundations and Trends® in Information Retrieval 225.
[36] See Annexes II and III AI Act.
[37] Philipp Hacker, Andreas Engel and Marco Mauer, 'Regulating ChatGPT and Other Large Generative AI Models', Proceedings of the 2023 ACM Conference on Fairness, Accountability, and Transparency (Association for Computing Machinery 2023) <https://dl.acm.org/doi/10.1145/3593013.3594067> accessed 17 July 2023.
[38] Philipp Bongartz, Sarah Langenstein and Rupprecht Podszun, 'The Digital Markets act: Moving from Competition Law to Regulation for Large Gatekeepers' (2021) 10 EuCML 60, 61.
[39] OJ L 186, 11.7.2019, p 57.
[40] OJ L 304, 22.11.2011, p. 64.
[41] OJ L 149, 11.6.2005, p. 22.
[42] Directive 2019/2161/EU of 27 November 2019 on better enforcement and modernisation of Union consumer protection laws, OJ L 328. 18.12.2019, p. 7.
[43] OJ L 119, 4.5.2016, p. 1.
[44] The term 'parameter' is poorly chosen, as in technical terms it is understood to mean the internal coefficients of the model and not the factors relevant to the decision, such as price, availability, etc. However, this seems to be meant in the case of the P2B Regulation and the CRD, see Recital 24 P2B Regulation and Recital 22 Omnibus Directive; see also Christian Alexander, 'Neue Transparenzanforderungen im Internet – Ergänzungen der UGP-RL durch den "New Deal for Consumers"' [2019] WRP 1235, marginal no. 30. Technically, the decision factors

must also disclose the reasons for the relative weighting of these parameters; search engines only have to disclose the relative weighting of the main parameters themselves. These requirements, while seemingly innocuous at first glance, are challenging concerning AI systems. For a long time, a dispute has raged about the extent to which current data protection law requires the disclosure of individual parameters (*features*) that the AI model analyses, and of their relative weighting.[45] This is relevant because general statements on the most important parameters determining the model's predictions are difficult to obtain or hardly possible for some advanced types of machine learning.[46] With support vector machines or artificial neural networks, in particular, techniques exist to explore, and only approximately, which features were decisive for an individual, concrete prediction (so-called *local* explanation).[47] It is technically much more difficult to determine those features that generally (i.e. concerning all decisions) are the most relevant ones (so-called *global* explanation).[48] However, Article 5 P2B Regulation arguably requires precisely this: the disclosure of general main parameters, i.e. global explanations of the AI model.[49] This requirement is not only a considerable legal innovation but also poses significant challenges for developers, especially when using artificial neural networks, which are often particularly potent.

### b) Consumer Rights Directive (CRD)

Similar considerations apply to the amendment of the Consumer Rights Directive (CRD). As with the P2B Regulation, the CRD is applicable regardless of the market power of the addressees, but sets obligations vis-à-vis consumers. According to the new Article 6a CRD, online marketplaces[50] must disclose the main parameters for rankings based on consumer search queries and their relative weighting. The provisions wording suggests an obligation to provide

---

are rather called 'features' (Ian Goodfellow, Yoshua Bengio and Aaron Courville, Deep Learning (MIT Press 2016), 3, 292 f).
[45] See for example Lea Katharina Kumkar and David Roth-Isigkeit, 'Erklärungspflichten bei automatisierten Datenverarbeitungen nach der DSGVO' [2020] JZ 277.
[46] Zachary C Lipton, 'The Mythos of Model Interpretability: In machine learning, the concept of interpretability is both important and slippery' (2018) 16 Queue 31.
[47] Scott M Lundberg and Su-In Lee, 'A unified approach to interpreting model predictions' (2017) 30 Advances in Neural Information Processing Systems 4765; Alejandro Barredo Arrieta et al., 'Explainable Artificial Intelligence (XAI): Concepts, taxonomies, opportunities and challenges toward responsible AI' (2020) 58 Information Fusion 82, 92 et seqq.
[48] Alejandro Barredo Arrieta et al., 'Explainable Artificial Intelligence (XAI): Concepts, taxonomies, opportunities and challenges toward responsible AI' (2020) 58 Information Fusion 82, 90; for approaches, see, e.g., Lapuschkin et al., 'Unmasking Clever Hans predictors and assessing what machines really learn' (2019) 10 Nature Communication 1; likewise, the calculation of an average of so-called Shapley values, which yield local feature relevance, is possible, cf. Scott M Lundberg and Su-In Lee, 'A unified approach to interpreting model predictions' (2017) 30 Advances in Neural Information Processing Systems 4765.
[49] Recital 24 P2B Regulation; Grochowski et al., 'Algorithmic Transparency and Explainability for EU Consumer Protection: Unwrapping the Regulatory Premises' (2021) 8 Critical Analysis of Law 43, 52; Bibal et al., 'Legal requirements on explainability in machine learning' (2021) 29 Artificial Intelligence & Law 149, 161; Philipp Hacker and Jan-Hendrik Passoth, 'Varieties of AI Explanations Under the Law. From the GDPR to the AIA, and Beyond' in Holzinger et al. (eds), xxAI – Beyond Explainable AI, International Workshop on Extending Explainable AI Beyond Deep Models and Classifiers (Springer 2022), 343, 364.
[50] According to the new Article 2(1)(n) of the UCP Directive, an online marketplace is a "service enabling consumers to conclude distance contracts with other traders or consumers through the use of software, including a website, part of a website or an application operated by or on behalf of the trader".

a local explanation, which would render explanations more feasible in case of advanced machine learning techniques, such as artificial neural networks, as seen. However, Recital 21 of the Omnibus Directive clearly states that the ranking-related transparency obligations of the CRD should mirror those of the P2B Regulation. The 23rd Recital of the Omnibus Directive also emphasizes that traders owe no disclosure in individual cases.[51] Therefore, the obligation to provide global explanations remains, raising the aforementioned implementation challenges in handling artificial neural networks.

### c) Unfair Commercial Practices Directive (UCPD)

The new Article 7(4a) of the UCPD accompanies the CRD disclosure obligation. It now qualifies the compulsory information on the main parameters laid down in Article 6a CRD as essential information in the sense of the prohibition of misleading information under unfair competition law. Hence, every violation of the CRD requirement automatically constitutes an unfair commercial practice, triggering, inter alia, legal action by competitors not foreseen under the CRD (Article 11(1) UCPD). In practice, these proceedings are among the most effective incentives for compliance by businesses with unfair competition law.

Simultaneously, Article 7(4a) UCPD applies to all entrepreneurs who enable a search for products of different suppliers.[52] Finally, according to point 11a of Annex I, it is considered unfair if payments or paid advertising used to achieve a higher ranking position are not clearly disclosed by the platform.[53] Such ranking categories must now be unambiguously identified as 'sponsored' or marked similarly.

Overall, the UCPD provisions do not merely repeat the CRD obligations; rather, they significantly raise the likelihood that the CRD obligations do not remain a paper tiger, but have a real effect in the digital economy.

### d) GDPR

Even though it predates the Omnibus Directive and the P2B Regulation, the GDPR has gone even further in establishing transparency as one of the main principles of data protection (Article 5(1)(a) GDPR). The principle is primarily operationalized through the rights to information (Article 12-14 GDPR) and access (Article 15 GDPR). For any automated decision making (including profiling), data controllers are required to disclose "meaningful information" about the logic involved as well as the significance and the envisaged consequences of such processing for the data subject (Article 13(2)(f), 14(2)(g), 15(1)(h) GDPR). As has been previously outlined, most forms of truly efficient, high quality AI-based rankings are currently created by analyzing either the previous behavior of the specific person to which the information is being presented, or the average consumer behavior in general. Both of these

---

[51] See also Christian Alexander, 'Neue Transparenzanforderungen im Internet – Ergänzungen der UGP-RL durch den "New Deal for Consumers"' [2019] WRP 1235 para 34.
[52] Christian Alexander, 'Neue Transparenzanforderungen im Internet – Ergänzungen der UGP-RL durch den "New Deal for Consumers"' [2019] WRP 1235 para 29.
[53] In addition, the terms and conditions for sponsored ranking must be disclosed to business clients, Article 5(3) P2B Regulation.

determinations require the processing of personal data, thereby falling under the scope of the GDPR.[54]

### i. Additional Requirements for Automated Decision Making

The existence of automated decision making on its own is not necessarily sufficient to trigger the outlined transparency requirement, however. Articles 13, 14 and 15 refer to Article 22(1) GDPR, which famously adds two additional requirements: the decision must be based *solely* on automated processing; and needs to produce legal or *similarly significant* effects for the individual. In some cases of AI-based ranking, controllers may by now have included human involvement to avoid scrutiny under these provisions.[55] Typical website or product rankings will, however, be built purely automatically by gatekeeper AI, raising the question of significant effects.

Under the Article 29 Data Protection Working Party Guidelines on Automated Individual Decision-Making and Profiling (WP Guidelines), decisions qualify if they have the potential to significantly affect the circumstances, behavior or choices of the individual, or if they have a prolonged or permanent impact.[56] Since empirical studies have evidenced that rankings do, in fact, have considerable relevance to consumer decisions,[57] the possibility of such an impact cannot be excluded a priori. While searches conducted on trivial matters might not impact us relevantly, there are certainly instances in which the type and order of information presented can have serious ramifications, because of the kind of product (e.g., insurance) or time sensitivity (e.g., information on poisons). Controllers might argue that such consequences are purely hypothetical. Still, if a misguided decision was, in fact, made based on a ranking, an individual could retroactively seek an explanation regarding the reasons (Article 15(1)(h) GDPR); and controllers reasonably expecting such effects need to proactively disclose the required information (Article 13(2)(f), 14(2)(g) GDPR). In our view, some ranking models, which are commonly used, should be considered significant due to a cumulative effect. If individuals use Google every day, in almost all areas of their lives, an overall significant impact can hardly be denied. In that sense gatekeeper AI would—barring human involvement—automatically fall under the definition of Article 22(1) GDPR.

### ii. Meaningful information

As a result, meaningful information must be provided on the logic involved. Like numerous terms in the GDPR, the formulation "meaningful information" is inherently vague, leaving vast room for (mis)interpretation. Since it is part of the transparency requirements, which are meant to provide to data subjects an understanding concerning the processing of their personal data, it should most likely be understood as in "meaningful to the data subject".[58] Hence, information about the "logic involved in the processing" should focus, at a minimum, on the general rationale behind the system phrased in a consumer-friendly way, rather than making available a specific algorithm or model.[59] Simultaneously, the controller would need to disclose enough information for the individual to form an understanding about the underlying logic and exercise their data subject rights under Article 15-22 GDPR, where needed.[60]

---

[54] See, e.g., Frederik Zuiderveen Borgesius, 'Personal data processing for behavioural targeting: which legal basis?' (2015) 5 International Data Privacy Law 163.
[55] Sandra Wachter, Brent Mittelstadt and Luciano Floridi, 'Why a Right to Explanation of Automated Decision-Making Does Not Exist in the General Data Protection Regulation' (2017) 7 IDPL 76, 88.
[56] Article 29 Data Protection Working Party, 'Guidelines on Automated individual decision-making and Profiling for the purposes of Regulation 2016/679' (2017) WP 251, 21.
[57] Raluca Ursu, 'The Power of Rankings: Quantifying the Effect of Rankings on Online Consumer Search and Purchase Decisions', (2018) 37 Market Science 530; Masha Derakhshan et al., 'Product Ranking on Online Platforms' (2022) 68 Management Science 4024.

While the wording in Articles 13, 14 and 15 GDPR is identical, the rights should still be viewed separately.[61] In their privacy policies, controllers must opt for *general* descriptions, taking into consideration the average consumer (global explanations); this, however, does arguably not apply to access requests. Here, the data subject has the option to specify what kind of information would interest them (e.g., storage period; feature relevance) and hence, by selecting specific information, to shape the very meaning of "meaningful". Whether this, however, includes a subjective right to an explanation regarding a specific decision made by the AI is still subject to heated debate.[62] As is well known, the idea of a specific justification is supported by Sentence 4 of Recital 72, which lists a right of the data subject "to obtain an explanation of the decision reached after such assessment and to challenge the decision" as one of the safeguards to be implemented. Also, the Article 29 WP, while confirming that the disclosure of the full algorithm would not be required, still underlines that the provided information should enable the individual "to understand the reasons for the decision",[63] further supporting the idea of individual (= local) explanations. While none of these interpretations are binding, they do indicate a general tendency and are in line with the purpose of the access right: to provide data subjects with enough information to understand and contest data processing by exercising their data subject rights. Arguably, in numerous scenarios, an effective contestation presupposes a local explanation.[64] Hence, a flexible approach will have to be taken by controllers, adapting the information provided to the circumstances of each case, as long as the matter has not been decided by the CJEU.

---

[58] Article 29 Data Protection Working Party, 'Guidelines on Automated individual decision-making and Profiling for the purposes of Regulation 2016/679' (2017) WP 251, 25; Andrew D Selbst and Julia Powles, 'Meaningful information and the right to explanation' (2017) 7 IDPL 233, 236; Bart Custers and Anna-Sophie Heijne, 'The right of access in automated decision-making: The scope of article 15(1)(h) GDPR in theory and practice' [2022] Computer Law & Security Review 46, 5.

[59] Gabriela Zanfir-Fortuna, 'Article 13 Information to be provided where personal data are collected from the data subject', in Christopher Kuner et al. (eds), The EU General Data Protection Regulation (GDPR): A Commentary (online edn, Oxford Academic 2020) 430.

[60] Article 29 Data Protection Working Party, 'Guidelines on Automated individual decision-making and Profiling for the purposes of Regulation 2016/679' (2017) WP 251, 25; Philipp Hacker and Jan-Hendrik Passoth, 'Varieties of AI Explanations Under the Law. From the GDPR to the AIA, and Beyond' in Holzinger et al. (eds), xxAI – Beyond Explainable AI, International Workshop on Extending Explainable AI Beyond Deep Models and Classifiers (Springer 2022), 343, 349; Selbst and Powles, (n 58) 236; Custers and Heijne, (n. 58) 5.

[61] Matthias Bäcker, 'Article 15' in Jürgen Kühling and Benedikt Buchner, Datenschutzgrundverordnung BDSG (3rd edn, CH Beck 2020) para 27; Peter Bräutigam and Florian Schmidt-Wudy, 'Das geplante Auskunft- und Herausgaberecht des Betroffenen nach Article 15 der EU-Datenschutzgrundverordnung', [2015] CR 56, 62. The opposite is argued in: Lea Katharina Kumkar and David Roth-Isigkeit, 'A Criterion-Based Approach to GDPR's Explanation Requirements for Automated Individual Decision-Making' (2021) 12 J Intell Prop Info Tech & Elec Com L 289, 296.

[62] Gianclaudio Malgieri and Giovanni Comandé, 'Why a Right to Legibility of Automated Decision-Making Exists in the General Data Protection Regulation', (2017) 7 IDPL 243; Wachter, Mittelstadt and Floridi (n 55), 76 – 99; Selbst and Powles, (n 58), 233 – 242. Maja Brkan, 'Do algorithms rule the world? Algorithmic decision-making and data protection in the framework of the GDPR and beyond' (February 28, 2018) 91, 110 -119.

[63] Article 29 Data Protection Working Party, (n. 56) 25.

### iii. Emerging case law

Indeed, while the CJEU has not decided on algorithmic transparency yet, cases are starting to emerge in Member State courts.[65] Most notably, in March 2021, the District Court of Amsterdam ruled in a case against the ridesharing company Ola, which operates a platform similar to Uber in the Netherlands.[66] The company had established an algorithmic model for automatically penalizing drivers if a ride was canceled or invalid. The drivers sued to enforce their right of access under Article 15(1)(h) GDPR. In its judgment, the court interpreted the clause in favor of the applicants and required the platform to explain the logic involved in the decisions. More specifically, it recurred on the Article 29 WP Guidelines, according to which meaningful information about the logic involved implies the disclosure of "criteria relied on in reaching the decision".[67] The court concluded that "Ola must communicate the main assessment criteria and their role in the automated decision to [the drivers], so that they can understand the criteria on the basis of which the decisions were taken and they are able to check the correctness and lawfulness of the data processing".[68]

In the view of the court, the GDPR access right implies an explanation of algorithmic decisions by feature relevance (main assessment criteria and role). The wording ("automated decision") and purpose of the judgment (correctness check) suggests that these explanations must be individually personalized. This means that, as suggested by our above analysis, local explanations of the individual decisions are required. As mentioned, such explanations can be furnished by post-hoc explanation models, such as SHAP, even in the case of highly complex, "black box" artificial neural networks.[69] Furthermore, the judgment rightly highlights that explanations must be adapted to the cognitive and educational background of data subjects.

The judgment was until recently on appeal with a positive outcome for the plaintiffs – the Court of Appeals predominantly upheld District Court's decision.[70] Regarding the requirements to comply with the right to be informed under Article 15(1)(h) GDPR, the appeals court clarified in paragraph 3.48, referring to the WP Guidelines, that the information must be provided in such a way that the data subject can decide with sufficient knowledge whether to exercise his or her rights guaranteed by the GDPR. This information includes, in particular, information about factors that were taken into account in the decision-making process and their respective "weighting" at an aggregate level. The information must also be complete enough for the data subject to understand the reasons for the decision. Though the court also says that it also follows from the guidelines that it does not have to be a complex explanation of the algorithms used or a presentation of the entire algorithm. However, it says that what is required is at least information about what factors and the weighting of those factors Ola used to arrive at the ratings at issue, as well as a statement of information necessary to understand why.

---

[64] See also sources cited in n. 60.
[65] See, e.g., (French) Conseil Constitutionnel, Décision n° 2020-834 QPC du 3 avril 2020, Parcoursup; (Dutch) Rechtbank Den Haag, Case C-09-550982-HA ZA 18-388, SyrRI, ECLI:NL:RBDHA:2020:1878; (Italian) Corte Suprema di Cassazione, Judgment of 25 May 2021, Case 14381/2021,
[66] District Court of Amsterdam, Case C /13/689705/HA RK 20-258, Ola, ECLI:NL:RBAMS:2021:1019 [Ola Judgment]; see also Raphaël Gellert, Marvin van Bekkum and Frederik Zuiderveen Borgesius, 'The Ola & Uber judgments: for the first time a court recognises a GDPR right to an explanation for algorithmic decision-making' (EU Law Analysis, 28 April 2021) <https://eulawanalysis.blogspot.com/2021/04/the-ola-uber-judgments-for-first-time.html/> accessed on 12 December 2022.
[67] Article 29 Data Protection Working Party, 'Guidelines on Automated individual decision-making and Profiling for the purposes of Regulation 2016/679' (2017) WP 251, 25.
[68] Ola Judgment, para. 4.52; translation according to Anton Ekker, 'Dutch court rules on data transparency for Uber and Ola drivers' (Ekker Blog) <https://ekker.legal/en/2021/03/13/dutch-Court-rules-on-data-transparency-for-uber-and-ola-drivers/> accessed on 12 December 2022.
[69] See n. 47.
[70] The judgement can be found here: https://uitspraken.rechtspraak.nl/#!/details?id=ECLI:NL:GHAMS:2023:804; an unofficial translation can be found here: https://5b88ae42-7f11-4060-85ff-4724bbfed648.usrfiles.com/ugd/5b88ae_de414334d89844bea61deaaebedfbbfe.pdf; see also Jakob Turner, 'Amsterdam Court Upholds Appeal in Algorithmic Decision-Making Test Case: Drivers v Uber and Ola' Fountain Court Blog (June 4, 2023), https://www.fountaincourt.co.uk/2023/04/amsterdam-court-upholds-appeal-in-algorithmic-decision-making-test-case-drivers-v-uber-and-ola/.

This underlines the importance of purpose-driven interpretation of Article 15(1)(h) GDPR whose obligations require information meaningful for the respective data subjects to enable them to exercise their respective rights. More specifically, in our view, explanations must conform to reasonable expectations of the data controller concerning the data subjects' understanding. In sum, Article 15(1)(h) GDPR demands explanations of automated decision-making systems, including AI, that are concrete, local, and adapted to the respective audience.

### iv. Large Generative AI Models

Also noteworthy in the context of reviewing the emerging case law is the temporary ban put on ChatGPT by the Italian Data Protection Authority. In March 2023 the *Garante per la protezione dei dati personali* imposed an immediate temporary limitation on the processing of Italian users' data by OpenAI.[71] While the service was quickly reinstated for Italian users, based on a list of measures deployed by the provider,[72] concerns with regards to the compliance of such services with the GDPR remain.[73]

Large generative AI models will not always create a ranking in the context of providing the user with answer. As has however been outlined above, the interaction with the chatbot will frequently lead to the generation of such.[74] The question of transparency in the context of large generative AI Models, such as ChatGPT is also not limited to the above considerations regarding the provisions of meaningful information about the logic involved in the automated decision-making process under Art. 22(1), but also effect general question of transparency in accordance with Art. 13 and 14 of the GDPR. On the one hand, the service provider is collecting personal data directly from the user, in the form of processing his or her inquiries and requests (Art. 13), while on the other hand personal data has likely been used for the training of the model itself, without any prior contact o the data subject (Art. 14). Whereas the provisions of relevant information concerning the processing of user and account related data, should be relatively easy, defining parameters such as the affected data subjects, the categories of data concerned, or the length of the processing operation in the context of AI training data will be virtually impossible. Based on that controllers might be inclined to use the exemption of "disproportionate effort" established in Art. 14(5)(b).[75] Should that be the case, the same Article would however still be require them to introduce "*appropriate measures to protect the data subject's rights and freedoms and legitimate interests*" and arguably perform a balancing test between the effort and the interest of the individual.[76] How this could be achieved by the provider and to what extent they would even be able to use the mentioned exemption, considering that their processing is not done for archiving purposes in the public interest, scientific or historical research purposes or statistical purposes is highly questionable. The general lack of transparency of those models is particularly problematic, when taking into account the identified risk of inversion attacks, leading to the reproduction of the used training data,[77] In that sense further steps from the relevant Data Protection Authorities would be welcome.

---

[71] See: Garante per la protezione dei dati personali, 'Intelligenza artificiale: il Garante blocca ChatGPT. Raccolta illecita di dati personali. Assenza di sistemi per la verifica dell'età dei minori' <https://www.garanteprivacy.it:443/home/docweb/-/docweb-display/docweb/9870847> accessed 17 July 2023.
[72] https://www.gpdp.it/home/docweb/-/docweb-display/docweb/9881490#english, accessed 10 May 2023.
[73] Philipp Hacker, Andreas Engel and Marco Mauer, 'Regulating ChatGPT and Other Large Generative AI Models', Proceedings of the 2023 ACM Conference on Fairness, Accountability, and Transparency (Association for Computing Machinery 2023) <https://dl.acm.org/doi/10.1145/3593013.3594067> accessed 17 July 2023; Julia Möller-Klapperich, `ChatGPT und Co. – aus der Perspektive der Rechtswissenschaft` (2023), 4 Neue Justiz 144; Goodfellow, I., Pouget-Abadie, J., Mirza, M., Xu, B., Warde-Farley, D., Ozair, S., Courville, A. and Bengio, Y., Generative adversarial networks. Communications of the ACM, 63, 11 (2020), 139-144; Press Release: LfDI informiert sich bei OpenAI, wie ChatGPT datenschutzrechtlich funktioniert, https://www.baden-wuerttemberg.datenschutz.de/lfdi-informiert-sich-bei-openai-wie-chatgpt-datenschutzrechtlich-funktioniert/, accessed 11 May 2023.
[74] See, e.g., Philipp Hacker, Andreas Engel and Marco Mauer, 'Regulating ChatGPT and Other Large Generative AI Models', Proceedings of the 2023 ACM Conference on Fairness, Accountability, and Transparency (Association for Computing Machinery 2023) <https://dl.acm.org/doi/10.1145/3593013.3594067> accessed 17 July 2023.

### e) The DMA

The recently enacted DMA adds to the plethora of transparency requirements under EU law. Like the P2B Regulation and the CRD, the DMA contains specific transparency requirements for rankings in its Article 6(5)(2): the gatekeeper must carry out the ranking based on transparent conditions. However, the concept of transparency is not further defined or elaborated upon. This raises the question of how the DMA's transparency obligation should be interpreted: Does it go beyond those of the P2B Regulation, does it fall short of them, or are they rather congruent? In our view, the latter interpretation is the most convincing.[78]

---

[75] Philipp Hacker, Andreas Engel and Marco Mauer, 'Regulating ChatGPT and Other Large Generative AI Models', Proceedings of the 2023 ACM Conference on Fairness, Accountability, and Transparency (Association for Computing Machinery 2023) <https://dl.acm.org/doi/10.1145/3593013.3594067> accessed 17 July 2023.

[76] Art. 29 WP, Guidelines on transparency under Regulation 2016/679, 17/EN WP260 rev.01, para.64.

[77] Ian Goodfellow and others, 'Generative Adversarial Networks' (2020) 63 Communications of the ACM 139; Nicholas Carlini and others, 'Extracting Training Data from Diffusion Models' (arXiv, 30 January 2023) <http://arxiv.org/abs/2301.13188> accessed 17 July 2023; Plant et. al, 'You are what you Write: Preserving Privacy in the Era of Large Language Models' (April 2022) <https://www.researchgate.net/publication/360079388_You_Are_What_You_Write_Preserving_Privacy_in_the_Era_of_Large_Language_Models> accessed 17 July 2023.

[78] See also Dennis Brouwer, 'Towards a ban of discriminatory rankings by digital gatekeepers? Reflections on the proposal for a Digital Markets Act' (Internet Policy Review, January 11, 2021) <https://policyreview.info/articles/news/towards-ban-discriminatory-rankings-digital-gatekeepers-reflections-proposal-digital> accessed 7 December 2022, according to which fairness was already to be understood as transparency in the sense of the P2B Regulation under the Commission draft.

To start with, nothing in the Recitals of the DMA suggests that specific types of transparency or explanation (e.g., local or global explanations; sensitivity lists or contrasting explanations[79]) are required under Article 6(5) DMA. Rather, Recital 52 DMA does not refer to the GDPR, but to the Commission's Guidelines on transparency of rankings under the P2B Regulation, which is intended to facilitate implementation and enforcement of Article 6(5) DMA. If, however, the P2B guidelines are supposed to provide guidance for the implementation of the DMA transparency provisions as well, both requirements need to be identical.

The controversies surrounding the transparency requirements for AI-based models in Article 13 AI Act[80] (see next section) and the Article 13-15 GDPR[81] show that various approaches are conceivable in factual and legal terms. A completely independent interpretation of the DMA transparency requirements, detached from the surveyed existing obligations, would entail significant legal uncertainty and practical implementation difficulties. Hence, overall, the DMA transparency requirement, in our view, mirrors the P2B requirement, and is not influenced by the–differently worded–GDPR rules just analyzed.

### f) AI Act

The AI Act was designed to foster trustworthy AI,[82] of which transparency is a key element.[83] Hence, one would have expected the AI Act to provide precise and meaningful requirements for transparency and explanations in AI systems that go beyond the status quo in general EU law and the generic reference in Article 6(5) DMA. Unfortunately, however, the opposite is the case.[84]

The key transparency requirement for high-risk AI systems is contained in Article 13 AI Act. According to this provision, some specific information needs to be disclosed, such as 'the level of accuracy, including its metrics, robustness and cybersecurity' as well a description of the input and training data and expected output.[85]

---

[79] See, e.g., the overview in Holzinger et al., 'xxAI - Beyond Explainable Artificial Intelligence' in: Holzinger et al. (eds), xxAI - Beyond Explainable AI, International Workshop on Extending Explainable AI Beyond Deep Models and Classifiers (Springer 2022) 13; Alejandro Barredo Arrieta et al., 'Explainable Artificial Intelligence (XAI): Concepts, taxonomies, opportunities and challenges toward responsible AI' (2020) 58 Information Fusion 82; ; for an account of their accuracy, see Andrés Alonso and José Manuel Carbó, 'Accuracy of Explanations of Machine Learning Models for Credit Decision' (2022) Banco de España Working Paper 2222.
[80] Philipp Hacker and Jan-Hendrik Passoth, 'Varieties of AI Explanations Under the Law. From the GDPR to the AIA, and Beyond' in Holzinger et al. (eds), xxAI – Beyond Explainable AI, International Workshop on Extending Explainable AI Beyond Deep Models and Classifiers (Springer 2022) 343, 358 et seqq.
[81] See only Sandra Wachter, Brent Mittelstadt and Luciano Floridi, 'Why a Right to Explanation of Automated Decision-Making Does Not Exist in the General Data Protection Regulation' (2017)7 IDPL 76; Andrew D Selbst and Julia Powles, 'Meaningful information and the right to explanation' (2017) 7 IDPL 233; Lea Katharina Kumkar and David Roth-Isigkeit, 'Erklärungspflichten bei automatisierten Datenverarbeitungen nach der DSGVO' [2020] JZ 277.
[82] Recitals 5 and 62 AI Act.
[83] Wolter Pieters, 'Explanation and trust: what to tell the user in security and AI?' (2011) 13 Ethics and Information Technology 53.
[84] See, e.g., Philipp Hacker and Jan-Hendrik Passoth, 'Varieties of AI Explanations Under the Law. From the GDPR to the AIA, and Beyond' in Holzinger et al. (eds), xxAI – Beyond Explainable AI, International Workshop on Extending Explainable AI Beyond Deep Models and Classifiers (Springer 2022), 343, 357 et seqq.
[85] See Article 13(3)(b)(ii) and (iv)-(vi) AI Act.

These provisions, however, are generic, even though they do go beyond the current disclosure framework in many respects. Regarding rankings, however, Article 11 in conjunction with Annex IV AI Act provides for more detailed regulation. For example, a description of the system architecture needs to be offered as well as an assessment of potentially discriminatory impacts, and information on the 'relevance of the different parameters' (Annex IV(2)(b) AI Act). Taken together, these provisions therefore establish a disclosure regime for fairness metrics (discriminatory impact) and global explanations of the model in terms of the relevance of its features, i.e., the factors that mainly contribute to its output, averaged over all cases.[86] While the explanation requirement matches the provisions of the P2B Regulation and the updated CRD and UCPD (see above), but also replicates its implementation problems, the novel fairness disclosure hints at the increased relevance of fairness in AI systems and rankings (see below, II.4.).

All these disclosure obligations improve the ability of business users and end users, as well as associations, NGOs, and public authorities, to gain insight into the ranking criteria of online intermediary services, search engines and other ranking providers. However, they do not formulate any further substantial fairness requirements for the arrangement of the elements of the ranking itself. Such substantive obligations may, however, be found in the GDPR and other parts of the DMA.

### 3. Accuracy: rectification of rankings

Another data subject right found in the GDPR, which is linked to the principle of accuracy (Article 5(d) GDPR), rather than transparency, but which also potentially affects AI-based ranking is the right to rectification enshrined in Article 16 GDPR. It provides that data subjects may obtain rectification of their personal data if they are inaccurate. This, of course, would apply to any data sets processed by gatekeepers, including those rankings are based on. What is unclear, however, is to what extent the right of rectification could apply to the ranking itself.

For example, the underlying information may have been correct, but the individual in question might still consider the *outcome* to be wrong. Imagine that a person frequently travels to London for work and spends significant time online researching opportunities to buy tickets for a football game by Chelsea London. She is simply intrigued by the new setup of the team after the change of ownership. As a result, whenever she searches for pullovers online, the first items on the ranking invariably include some merchandise from Chelsea. This, however, is perceived as deeply offending by the London traveler as she is, in reality, an inveterate fan of Bayern Munich, but never wears any football merchandise. She is not opposed to personalized rankings, but would like them to be accurate. Does she then have a right for these personalized rankings to be rectified to better match her preferences?

---

[86] See, e.g., Sebastian Lapuschkin et al., 'Unmasking Clever Hans predictors and assessing what machines really learn' (2019) 10 Nature Communication 1.

### a) Rankings as Personal Data

Since the right to rectification only applies to personal data, the first question is whether the ranking would even fall under that definition. As per Article 4(1) GDPR, the term includes *any information relating to an identified or identifiable natural person*. Non-personalized rankings in the sense of prominence given to goods or services offered online or the relevance of search results (Article 2(22) DMA) do not generally fall under that definition. While rankings may *contain* personal data (e.g., if one googles a person), the ranking model itself solely constitutes a grouping of the relevant elements, based on the preferences of the average user.[87]

To achieve more accurate results, however, providers mostly rely on personalized recommendations, based on criteria such as preferences disclosed by the user, previous activities or social connections.[88] The presented outcome is therefore created *specifically* for that individual. To be personal data, though, it also has to *relate* to that person, whereas in this case it is rather *presumed to relate*, according to the model. While the simple fact of potential inaccuracy is not a reason for exclusion (after all, most personal data can be incorrect as well),[89] the legal status of inferences is questionable.[90]

A wide concept of personal data, including personalized rankings, seems to be supported by the general trend in interpreting that concept. The Article 29 WP, in its Opinion on the Concept of Personal Data, suggests that information can relate to a natural person based on content, purpose, or result.[91] That same broad understanding of the term was supported by the CJEU in *Nowak*.[92] Based on this categorization, the element of 'content' is present where information is 'about' a person; 'purpose' where the data is used to evaluate that individual; and 'result' where it is likely to impact their rights and interests.[93]

---

[87] Liu Tie-Yan, 'Learning to Rank for Information Retrieval' (2009) 3 Foundations and Trends in Information Retrieval 225.; Thorsten Joachims et al., 'Accurately interpreting clickthrough data as implicit feedback' in Proceedings of the 28th Annual International ACM SIGIR Conference on Research and Development in Information Retrieval (2005) 154; Thorsten Joachims et al., 'Accurately interpreting clickthrough data as implicit feedback' in Proceedings of the 28th Annual International ACM SIGIR Conference on Research and Development in Information Retrieval (2005) 154.

[88] G Adomavicius and A Tuzhilin, 'Toward the next generation of recommender systems: a survey of the state-of-the-art and possible extensions' in (2005) 17 IEEE Transactions on Knowledge and Data Engineering 734; Gábor Takács and Domonkos Tikk, 'Alternating least squares for personalized ranking' in RecSys '12: Proceedings of the sixth ACM conference on Recommender systems (Association for Computing Machinery 2012) 83.; Tong Zhao, Julian McAuley and Irwin King, 'Leveraging Social Connections to Improve Personalized Ranking for Collaborative Filtering' in CIKM'14: Proceedings of the 23rd ACM International Conference on   Conference on Information and Knowledge Management (Association for Computing Machinery 2014) 261.

[89] The fact that personal data can be incorrect is explicitly recognised by the GDPR, through the introduction of the right to rectification (see: Article 29 Working Party, 'Opinion 4/2007 on the concept of personal data', WP 136 (01248/07/EN) 6 <https://ec.europa.eu/justice/article-29/documentation/opinion-recommendation/files/2007/wp136_en.pdf > accessed: 17. November 2022, also arguably opinions do not need to be held to the standards of accuracy at all, see: Dara Hallinan, Frederik Zuiderveen Borgesius, 'Opinions can be incorrect (in our opinion)! On data protection law's accuracy principle' (2020) 10 International Data Privacy Law 1.

[90] Sandra Wachter and Brent Mittelstadt, 'A Right to Reasonable Inferences: Re-thinking Data Protection Law in the Age of Big Data and AI' (2019) 2 Colum Bus L Rev 494.

[91] Article 29 Working Party, 'Opinion 4/2007 on the concept of personal data', WP 136 (01248/07/EN) 10 <https://ec.europa.eu/justice/article-29/documentation/opinion-recommendation/files/2007/wp136_en.pdf    > accessed: 17. November 2022.

[92] Case C-434/16 Nowak v Data Protection Commissioner, ECLI:EU:C:2017:994, para. 35.

Arguably, a ranking is not primarily used for evaluation of the data subject; if anything, it *constitutes* an evaluation. As already outlined in discussing its significant effect, however, the potential impact on the individual cannot be denied. Also, the entire purpose of the ranking is, in fact, the provision of information most relevant, or otherwise optimized, for a specific person. Additionally, it might be considered whether, in fact, the ranking does not constitute new information *about* that person. This at least is the opinion of the Article 29 WP Opinion regarding profiling, which is described as "creating derived or inferred data about individuals– 'new' personal data that has not been provided directly by the data subjects themselves".[94] In our view, personalized rankings, therefore, do generally constitute personal data.

### b)    Rectification

This raises the question of the ranking's rectifiability. While the Article 29 WP Opinion considered the right to apply to profiles and scores as well,[95] the practical implementation poses significant challenges. After all, the ranking does not exist independently and objectively, but stems from the individual's personal data, as analyzed by the algorithmic model. If the accuracy of the underlying information is not in question, one element which could be challenged is the ranking method itself. The applied method, however, is generally the same for each data subject and thus cannot constitute personal data.

This differentiation between the underlying data and the analytical method used is also supported by the CJEU's ruling on the joint cases of *YS and M. and S. vs. Minister door Immigratie, Integratie en Asiel*.[96] Here, the Court decided that while the applicant's data on a resident permit, which constitutes the basis for the decision, does constitute personal data, the legal analysis itself does not.[97] The same logic could be applied to rankings: whereas the user's data constitutes personal data, as well as the outcome, the used ranking algorithm (the analysis) does not and is, therefore, not subject to rectification.

If an individual seeks to obtain a rectification of such a ranking, they would be left with the option of requesting rectification for the other two elements (analyzed data, outcome). On the one hand, the direct revision of the outcome could be demanded. In that case, the data subject could submit a manually created, "correct" ranking to the controller, or indicate which items are incorrectly highlighted in the ranking. The controller would then need to revise the ranking algorithm for the data subject to exclude such items. The user would constitute, so to speak, an additional supervision instance for training the model.

Another option would be to revise the underlying data. If its accuracy is not in question, however, that would in essence leave the option of completion as Article 17 also provides individuals with the right to have "incomplete personal data completed, including by means of providing a supplementary statement". Both solutions hold the potential to correct personalized rankings, but practical problems in challenging incorrect and implementing novel rankings are likely to persist.

---

[93] Ibid, 10 – 11.
[94] Article 29 Data Protection Working Party, (n. 56), 9.
[95] Article 29 Data Protection Working Party, (n. 56),   18.
[96] Case C-141/12 YS v Minister voor Immigratie, ECLI:EU:C:2014:2081.
[97] ibid. para. 45 – 46.

### c)    Article 15 AIA Act and the AILD

Recently, the legislator has added another instrument for contesting inaccurate AI decisions of all sorts. Article 15(1) AI Act holds that high-risk systems need to achieve an appropriate level of accuracy.[98] While the AI Act itself does not contain a private enforcement mechanism, private rights of action complementing the AI Act are provided by the new AI liability package unveiled by the Commission in September 2022.[99] Hence, if the AI developer violates Article 15(1) AI Act, the causal link between this fault and the AI output is presumed, for tort law claims noted in Member State law, according to Article 4(2) AILD Proposal. While this provision is obsolete in the case of a breach of the performance requirements–as the incorrect AI output *is* the fault[100]–, the data subject can nevertheless claim damages based on national tort law. She may also obtain relevant evidence from the AI provider to the extent that this is necessary to prove the claim (Article 3 AILD Proposal). Moreover, the updated PLD can be invoked, under which a violation of the AI Act will probably indicate the defectiveness of the AI product.[101]

The damage would consist, inter alia, in the provision of an incorrect ranking; remedying the damage would include providing a correct ranking, or at least one with an appropriate level of accuracy in the sense of Article 15(1) AI Act. This includes not only cases of personalized rankings but also non-personalized ones, such as candidate lists outputted by hiring algorithms. However, unfortunately, such types of damage are not covered by the PLD framework (Article 4(6) PLD Proposal). Hence, such rectification can only be sought under national tort law regimes,[102] unless the PLD Proposal is changed to incorporate certain types of pure economic or immaterial damage.[103]

### 4.    Fair Rankings

The final, and perhaps most comprehensive pillar of ranking regulation beyond transparency and accuracy concerns fairness, i.e., non-discrimination in various guises.

---

[98] Technically speaking, the correct term would be 'performance', with accuracy being only one of several relevant performance measures, see Andreas Lindholm et al., Machine Learning - A First Course for Engineers and Scientists (CUP 2022) 88.
[99] See, e.g., Philipp Hacker, 'The European AI Liability Directives--Critique of a Half-Hearted Approach and Lessons for the Future' (2022) arXiv preprint arXiv:221113960, 5 et seqq.
[100] Cf. Philipp Hacker, 'The European AI Liability Directives--Critique of a Half-Hearted Approach and Lessons for the Future' (2022) arXiv preprint arXiv:221113960, 34, 37.
[101] See Article 6 PLD Proposal and Philipp Hacker, 'The European AI Liability Directives--Critique of a Half-Hearted Approach and Lessons for the Future' (2022) arXiv preprint arXiv:221113960, 22, 53.
[102] European Commission, 'Questions and answers on the revision of the Product Liability Directive' (QANDA/22/5791, 2022), under 9.
[103] For an argument in this vein, see Philipp Hacker, 'The European AI Liability Directives--Critique of a Half-Hearted Approach and Lessons for the Future' (2022) arXiv preprint arXiv:221113960, 44.

### a)     Lessons from Competition Law

On November 10, 2021, the CJEU dismissed an action by Google challenging a €2.4 billion fine imposed by the Commission and confirmed the allegation that Google had intended to weaken the competitors' market position by favoring results from its own shopping comparison service over those of competitors.[104] Similarly, the Commission has been conducting several proceedings against Amazon since November 2020, among other things to examine whether the company gives preferential treatments to its own offers and offers from sellers using Amazon's logistics and shipping services.[105]

These cases exemplify the business models of many online platforms. On the one hand, they attempt to establish a business relationship between their business users, i.e., merchants and end users (*matching*). On the other hand, the companies behind the platforms sometimes also place their own production on these platforms, which they sell to the end users and thus enter direct competition with the business users (*dual mode*).[106] As a result, the platforms can shape the rankings to their advantage, disadvantaging their competitors and thus distorting competition.[107] Under Article 102 TFEU, such self-preferential treatment is (correctly) classified as exclusionary conduct,[108] as it leads to palpable distortions of competition and welfare losses.[109]

Article 102 TFEU is the central provision in proceedings such as those described above, which prohibits the abuse just described and represents the oldest and perhaps best-known rule on fairness of rankings in e-commerce. Thus, current competition law already prevents companies from favoring their products over competitors' ones on their own platforms. This prohibition of self-preference is a core component of substantial fairness regulations in rankings.[110]

However, the first prerequisite for the applicability of Article 102 TFEU is proof of a dominant position.[111] These proceedings necessitate an encompassing, economically oriented ex-post review, are often time-consuming, and thus generally come too late, particularly for smaller competitors.[112] For example, the proceedings described above did not prevent Google from significantly improving its economic position[113]–a result that does not effectively counteract the abuse of a dominant position. The DMA addresses, among other things, precisely this time lag of competition law (Recital 5 DMA).

---

[104] CJEU, Case T-612/17, Google Shopping, ECLI:EU:T:2021:763; on the procedure, see, for example, Andrea Lohse, 'Marktmissbrauch durch Internetplattformen?' (2018) 182 ZHR 321, 348-53.
[105] European Commission, Procedure AT.40703 (Amazon Buy Box).
[106] Andrei Hagiu, Tat-How Teh and Julian Wright, 'Should platforms be allowed to sell on their own marketplaces?' (2022) 53 The RAND Journal of Economics 297; Philipp Bongartz, Sarah Langenstein and Rupprecht Podszun, 'The Digital Markets act: Moving from Competition Law to Regulation for Large Gatekeepers' (2021) 10 EuCML 60; Jorge Padilla, Joe Perkins and Salvatore Piccolo, 'Self-Preferencing in Markets with Vertically Integrated Gatekeeper Platforms' (2022) 70 The Journal of Industrial Economics 371.
[107] Inga Graef, 'Differentiated Treatment in Platform-to-Business Relations: EU Competition Law and Economic Dependence' (2019) 38 Yearbook of European Law 453.
[108] Inga Graef, 'Differentiated Treatment in Platform-to-Business Relations: EU Competition Law and Economic Dependence' (2019) 38 Yearbook of European Law 453.
[109] Andrei Hagiu, Tat-How Teh and Julian Wright, 'Should platforms be allowed to sell on their own marketplaces?' (2022) 53 The RAND Journal of Economics 297, 300.
[110] Philipp Bongartz, Sarah Langenstein and Rupprecht Podszun, 'The Digital Markets act: Moving from Competition Law to Regulation for Large Gatekeepers' (2021) 10 EuCML 60, 62.
[111] This can lead to a problematic market definition, particularly in the case of digital companies; see, for example, Bundeskartellamt, decision of February 6,2019, B6-22/16, para. 166 et seq.
[112] Philip Marsden, 'Google Shopping for the Empress's New Clothes –When a Remedy Isn't a Remedy (and How to Fix it)' (2020) 11 Journal of European Competition Law & Practice 553; Digital Competition Expert Panel, 'Unlocking Digital Competition' (Report, 2019) para. 2.46.

### b) The Digital Markets Act

With Article 6(5), the DMA has now introduced a central provision for regulating AI-based rankings–a provision that is probably the most far-reaching one for regulating AI applications by gatekeepers. The provision comprises three components: first, the gatekeeper may not favor its products or services over those of third parties in a ranking (Article 6(5)(1) DMA). Second, the ranking must be transparent in general (see above); and third, fair and non-discriminatory (Article 6(5)(2) DMA).

#### i. Prohibition of self-preference

Like general competition law, the DMA addresses the prohibition of self-preference. However, Article 6(5)(1) DMA draws consequences from the weaknesses of enforcement of Article 102 TFEU outlined above:[114] unlike there, sanctions for self-preferential treatment under Article 6(5) DMA are not tied to an (elaborate) procedure to prove the dominant position, the anti-competitive effects, and the market definition; and justification by positive welfare effects is also impossible.[115] This sets the stage for more effective enforcement and, as a result, compliance by deterrence. Further, compared to the Commission's proposal, the final version of the DMA has extended this prohibition of self-preference to indexing and crawling.

#### ii. Non-discrimination

Next to the prohibition of self-preference and the transparency requirement, Article 6(5)(2) DMA stipulates that rankings must be based on "fair and non-discriminatory conditions".[116] General non-discrimination provisions for rankings, such as these, have already been discussed in the antitrust literature[117] and offer a point of reference the DMA clause as well.

##### (1) Relationship to FRAND conditions

Fairness and non-discrimination, in Article 6(5) DMA, cannot be reduced to transparency.[118] Rather, the wording of Article 6(5)(2) DMA explicitly distinguishes between transparency, fairness and non-discrimination. However, the fairness and non-discrimination obligation can, from a systematic perspective, be linked to criteria known from general antitrust law, according to which dominant undertakings must treat their corporate customers in fair, reasonable and non-discriminatory ways (in short: FRAND) in certain areas.[119] This obligation typically follows from Article 102 TFEU.[120] It is particularly relevant in intellectual property law, where owners of, for example, standard-essential patents, i.e. intellectual property rights that are essential for market access, must observe FRAND criteria when licensing.[121]

---

[113] Cf. Torsten J Gerpott, 'Das Gesetz über digitale Märkte nach den Trilog-Verhandlungen' [2022] CR 409, according to which Google's share value has increased by a factor of five during the almost 11-year duration of the proceedings (citing Macrotrends, 'Alphabet Market Cap 2010-2021'<https://www.macrotrends.net/stocks/charts/GOOGL/alphabet/market-cap> accessed September 12, 2022, for stock market value data).
[114] See II.4.a).
[115] Eifert et al. (n. 15), 1003 f.
[116] Brouwer, (n. 78).
[117] Inga Graef, 'Differentiated Treatment in Platform-to-Business Relations: EU Competition Law and Economic Dependence' (2019) 38 Yearbook of European Law 453, 463 f.
[118] But see Brouwer (n 78).

In the DMA itself, the FRAND formulation appears explicitly in Article 6(11), according to which gatekeepers must grant search engine operators access to ranking, search, click and display data of search results on FRAND terms.[122] Article 6(12) DMA requires FRAND conditions for access to app stores, among others. It is striking, however, that the wording of Article 6(5) DMA incorporates the FRAND language but, in contrast to Article 6(11) and (12) DMA, removes the attribute of "reasonableness". In the FRAND context, this criterion regularly refers to the access or licence conditions, particularly the amount of the license fee.[123] Rankings created by gatekeepers may, of course, include items free of charge, so that no fee is paid by business users. However, the reasonableness control refers not only to the amount of the fee, but also to other access conditions.[124] Indeed, special sections of the ranking are often reserved for paid content ("sponsored content").[125] In this respect, there would undoubtedly have been room for a reasonableness test. Ultimately, the decision to exclude a reasonableness review can only be explained by the reluctance of the EU legislator to trigger a complex control of conditions and, above all, prices in the context of rankings.[126] Incidentally, the transparency obligations described above do apply to sponsored rankings, as seen.

Hence, all rankings, regardless of whether they are sponsored, must be fair and non-discriminatory (FAND) under the DMA. However, the fairness element typically has no independent meaning within FRAND law.[127] Admittedly, Article 12(5) DMA lists practices that must each be considered as limiting the contestability of platform services or as unfair. The second and third practice mentioned there[128] lack any reference to rankings–unlike the third practice, which addresses barriers to market entry.[129] However, the wording of this variant clearly refers to the contestability, and thus not to the fairness, of platform services. Hence, the FRAND jurisprudence and the DMA do not offer any support for a separate fairness criterion for rankings. Therefore, rankings under the DMA need only be non-discriminatory–which may be difficult enough, as we shall presently see.

---

[119] See only Jonas Block and Benjamin Rätz, 'The FRAND offer - attempt at an international definition' [2019] GRUR 797.; Jean-Sébastien Borghetti, Igor Nikolic and Nicolas Petit, 'FRAND licensing levels under EU law' (2021) 17 European Competition Journal 205; J Gregory Sidak, 'The Meaning of FRAND, Part I: Royalties' (2013) 9 Journal of Competition Law and Economics 931; Matthew Heim and Igor Nikolic, 'A FRAND Regime for Dominant Digital Platforms' (2019) 10 Journal of Intellectual Property, Information Technology and Electronic Commerce Law 38.
[120] See only Jonas Block and Benjamin Rätz, 'The FRAND offer - attempt at an international definition' [2019] GRUR 797.; Jean-Sébastien Borghetti, Igor Nikolic and Nicolas Petit, 'FRAND licensing levels under EU law' (2021) 17 European Competition Journal 205; J Gregory Sidak, 'The Meaning of FRAND, Part I: Royalties' (2013) 9 Journal of Competition Law and Economics 931; Matthew Heim and Igor Nikolic, 'A FRAND Regime for Dominant Digital Platforms' (2019) 10 Journal of Intellectual Property, Information Technology and Electronic Commerce Law 38.
[121] See, for example, Ann, Patentrecht, (8th ed, CH Beck 2022) § 43 marginal no. 35 et seqq.
[122] Peter Georg Picht and Heiko Richter, 'EU Digital Regulation 2022: Data Desiderata' [2022] GRUR International 395.
[123] J Gregory Sidak, 'The Meaning of FRAND, Part I: Royalties' (2013) 9 Journal of Competition Law and Economics 931.).
[124] Ann, Patentrecht, (8th ed, CH Beck 2022) § 43 marginal no. 36.
[125] Christoph Busch, 'Mehr Fairness und Transparenz in der Plattformökonomie? Die neue P2B-Verordnung im Überblick' [2019] GRUR 788.
[126] See, for example, Ann, Patentrecht, (8th ed, CH Beck 2022) § 43 marginal no. 39 ff on the various calculation approaches.
[127] J Gregory Sidak, 'The Meaning of FRAND, Part I: Royalties' (2013) 9 Journal of Competition Law and Economics 931, 968; Brouwer (fn. 78) would like to load fairness in the ranking context, albeit still on the basis of the original Commission draft, with the transparency rules of the P2B Regulation. This is now prohibited, as transparency is an independent criterion.

### (2) Non-discriminatory rankings: between non-discrimination and competition law

Platforms regularly create rankings based on specific product attributes. Importantly, the non-discrimination requirement arguably refers to these same product attributes.[130] The FAND condition in Article 6(5) DMA is not limited to mere non-discriminatory *access* to the ranking results. This follows from a comparison with Article 6(11)(1) DMA: here, FRAND conditions are restricted to access to certain data. *E contrario*, the FAND clause in Article 6(5) DMA must encompass more than access, i.e., the order of the ranking itself.

The link to the FRAND literature provides a starting point for a delimiting the meaning of non-discrimination with respect to DMA rankings. As in the case of FRAND conditions, what exactly constitutes a protected attribute under Article 6(5) DMA will probably depend to a certain degree on the individual case. However, certain guidelines can be established based on (non-exhaustive) groups of cases: In the first group, discrimination refers to the anthropocentric attributes protected by classical non-discrimination law,[131] such as gender, religion, racial or ethnic origin, and nationality. This group of attributes also corresponds to the ongoing research effort in computer science on fair, i.e., non-discriminatory rankings.[132]

### (a) Coherence between non-discrimination and competition law

Within this first group of cases, one may differentiate again: The list may directly rank persons or groups of persons (such as the providers of services or works, or users listed in social media feeds). In this case, significantly, non-discrimination doctrine should guide the interpretation of Article 6(5) DMA.[133] This corresponds to a coherent interpretation of EU law that the CJEU

---

[128] This concerns access to essential input data and equivalence violations in the contractual structure.
[129] Article 12 para. 5 lit. b (i) DMA.
[130] Cabral and others, 'The EU Digital Markets Act', (Joint Research Centre, JRC122910, Publications Office of the European Union 2021) 13.
[131] See, e.g., Council Directive 2000/43/EC of 29 June 2000 implementing the principle of equal treatment between persons irrespective of racial or ethnic origin [Race Equality Directive]; Council Directive 2000/78/EC of 27 November 2000 establishing a general framework for equal treatment in employment and occupation [Framework Directive]; Council Directive 2000/78/EC of 27 November 2000 establishing a general framework for equal treatment in employment and occupation; Council Directive 2004/113/EC of 13 December 2004 implementing the principle of equal treatment between men and women in the access to and supply of goods and services.
[132] See, for example, the overview in Abolfazl Asudeh and others, 'Designing Fair Ranking Schemes' in SIGMOD '19: Proceedings of the 2019 International Conference on Management of Data (Association for Computing Machinery 2019) 1259; Ashudeep Singh and Thorsten Joachims, 'Fairness of Exposure in Rankings' in Proceedings of the 24th ACM SIGKDD International Conference on Knowledge Discovery & Data Mining (Association for Computing Machinery 2018) 2219.; Meike Zehlike, Ke Yang and Julia Stoyanovich, 'Fairness in Ranking, Part II: Learning-to-Rank and Recommender Systems' (2022) 55 ACM Computing Surveys, Article 117.
[133] See, e.g., Sandra Wachter, 'The Theory of Artificial Immutability: Protecting Algorithmic Groups Under Anti-Discrimination Law' (2022) arXiv preprint arXiv:220501166; Sandra Wachter, Brent Mittelstadt and Chris Russell, 'Why fairness cannot be automated: Bridging the gap between EU non-discrimination law and AI' (2021)

has repeatedly demanded even across legal fields in cases such as *Pereničová and Perenič*,[134] *Bankia*,[135] and *Pillar Securitisation*.[136] The key take-away from these cases is that a concept, such as non-discrimination, transplanted from one area of EU law to another should be interpreted coherently, while acknowledging the idiosyncrasies of the respective fields. In the words of AG Trstenjak, "what is needed is a coherent interpretation of the relevant rules of law so as to avoid conflicting assessments".[137] Therefore, in our view, a finding of discrimination under non-discrimination law does not automatically imply a violation of Article 6(5) DMA, but offers a strong indication of the use of discriminatory conditions according to the DMA unless DMA-specific justifications can be found.[138]

Hence, if persons are ranked, direct or indirect discrimination may occur in the ranking.[139] For example, if ranking parameters include attributes of these persons or groups of persons protected by classical non-discrimination law, such as gender, this will generally constitute direct discrimination, unless a justification applies. Such a justification may be found in traditional non-discrimination law or in an implicit, DMA-specific justification modeled on the FRAND literature and jurisprudence. For example, restricting the ranking to female persons may be justified if the female gender constitutes a genuine and determining occupational requirement.[140] This implies that the protected attribute is essential for the task at hand.[141] search for justification fails, for example, if a platform intermediates jobs for merely historically female jobs (e.g., cleaning) which are now performed by persons of any gender.

In our view, another subcategory of this group of cases concerns the ranking of goods or companies which is influenced by traditionally protected attributes of persons related to these items. For example, rankings of companies and their products may generally not be influenced by the religion or gender of their CEOs (or other company members). Importantly, even customer preferences do not provide a justification, according to the CJEU, in such cases.[142]

---

41 Computer Law & Security Review 105567; Frederik J Zuiderveen Borgesius, 'Strengthening legal protection against discrimination by algorithms and artificial intelligence' (2020) 24 The International Journal of Human Rights 1572; Meike Zehlike, Philipp Hacker and Emil Wiedemann, 'Matching code and law: achieving algorithmic fairness with optimal transport' (2020) 34 Data Mining and Knowledge Discovery 163; Frederik Zuiderveen Borgesius, 'Discrimination, artificial intelligence, and algorithmic decision-making' (Council of Europe, Directorate General of Democracy 2018); Philipp Hacker, 'Teaching fairness to artificial intelligence: existing and novel strategies against algorithmic discrimination under EU law' (2018) 55 Common Market Law Review 1143.
[134] CJEU, Case C-453/10, Pereničová and Perenič, ECLI:EU:C:2012:144.
[135] CJEU, Case C-109/17, Bankia, ECLI:EU:C:2018:201.
[136] CJEU, Case C-694/17, Pillar Securitisation, ECLI:EU:C:2019:44, para. 35.
[137] AG Trstenjak, Case C-453/10, Pereničová und Perenič, ECLI:EU:C:2011:788, para. 90 (discussing legal acts relating to EU consumer law).
[138] See also Philipp Hacker, Datenprivatrecht (Mohr Siebeck 2020) 335 et seqq.
[139] See, e.g., Jeremias Adams-Prassl, Reuben Binns and Aislinn Kelly-Lyth, 'Directly Discriminatory Algorithms' (2022) The Modern Law Review.
[140] See, e.g., Art. 4(1) of the Race Equality Directive; Art. 4(1) of the Framework Directive; Art. 14(2) of the recast Gender Equality Directive 2006/54/EC; CJEU, Case C-229/08, Wolf, ECLI EU:C:2010:3, para. 35.
[141] Philipp Hacker, 'Teaching fairness to artificial intelligence: existing and novel strategies against algorithmic discrimination under EU law' (2018) 55 Common Market Law Review 1143, 1166.
[142] CJEU, Case C-188/15, Bougnaoui, EU:C:2017:204, para. 40; Joined Cases C-804/18 and C-341/19, Wabe and Müller, ECLI EU:C:2021:594, para. 65; see also Erica Howard, 'Headscarves and the CJEU: Protecting fundamental rights or pandering to prejudice' (2021) 28 Maastricht Journal of European and Comparative Law 648, 255 et seqq.; Elke Cloots, 'Safe Harbour or Open Sea for Corporate Headscarf bans? Achbita and Bougnaoui', (2018) 55 Common Market Law Review 589, 613.

### (b) Beyond non-discrimination law

The second group of cases concerns discrimination as it is understood in competition law. For example, the ranking position of a product may depend on whether the offering company has concluded an exclusivity agreement with the gatekeeper. Such a ranking also constitutes discrimination within the meaning of Article 6(5) DMA.[143]

This example indicates that the concept of discrimination in Article 6(5) DMA does not merely correspond to that of classical non-discrimination law, but must go beyond it. Products do not exhibit any of the traditionally protected anthropocentric attributes, such as gender. Clearly, it cannot be considered discrimination if the query itself, in a legitimate way, restricts the search to goods particularly valuable to certain protected groups, such as female or male shoes. Limiting the scope of the prohibition of discrimination in Article 6(5) DMA to traditionally protected attributes, however, would not meet the aim of the DMA, and in particular the ranking provisions, to ensure the contestability of markets.[144] Hence, a more comprehensive understanding of the prohibition of discrimination is needed. Indeed, it may link back to the requirements for entities covered by FRAND obligations: these may not treat similarly situated business users differently without justification.[145]

At the outset, this implies that the covered entities must offer the same conditions for comparable circumstances.[146] However, the purpose of a ranking is precisely to *distinguish* comparable products or services from each other to facilitate the customers' selection decision. To award the same ranking position twice would contradict this purpose. Hence, freedom from discrimination in the sense of Article 6(5) DMA means that distinctions must be objectively justifiable.[147] In our view, this entails that the ranking must be based on criteria that are relevant for the comparison of ranked products to ultimately facilitate the economic decision of consumers. If, for example, the ranking is based on whether business users use the logistics channels of the gatekeeper ("Fulfilment by Amazon"), the ranking constitutes prohibited discrimination that can only be justified in exceptional cases, according to the criteria just mentioned.

### iii. Technical possibilities and difficulties

The legal requirements just described also need to be technically implemented in the ranking models. The prohibition of self-preference, the transparency and the non-discrimination requirements must be operationalized so that the algorithm creating the ranking takes these into account and outputs a ranking that meets the requirements.

---

[143] Cabral and others, 'The EU Digital Markets Act', (Joint Research Centre, JRC122910, Publications Office of the European Union 2021) 13.
[144] In particular the contestability of ranked products and services, cf. Recital 7, 11, and 51 DMA.
[145] See, e.g., District Court of Düsseldorf, [2018] GRUR-RS 37930, ECLI:DE:LGD:2018:1212.4B.O4.17.00, para 202; CJEU, Case C-313/04, Egenberger, ECLI:EU:C:2008:728, para 33; Unwired Planet v Huawei [2018] EWCA Civ 2344, para 162; European Commission, 'Setting Out the EU Approach to Standard Essential Patents' (Communication) COM(2017) 712 final, 7; see also Brouwer (n. 78); Article 102(c) TFEU.
[146] Unwired Planet v Huawei [2018] EWCA Civ 2344, para 162 et seq.
[147] See also Brouwer (n. 78); Unwired Planet v Huawei [2018] EWCA Civ 2344, paras 169-170.

While implementing the prohibition of self-preference appears rather unproblematic,[148] the first major difficulties arise in the operationalization of the transparency obligations. This is due to the lack of the congruence and coordination of the transparency requirements under the DMA/P2B Regulation on the one hand and the GDPR as well as the AI Act on the other. Finally, technical challenges arise in implementing non-discrimination. To create a non-discriminatory AI-based ranking, the model must be audited for its impact on protected groups (e.g., gender and religion). A comprehensive exploration of algorithmic fairness techniques already exists in computer science.[149] If an unjustified unequal treatment is detected, adequately chosen constraints from the algorithmic fairness literature ensure that at least a similar number of persons or products of each of the protected classes are also represented in the upper part of the ranking.[150]

However, classical non-discrimination law defines only a finite, low number of protected attributes. In the case of Article 6(5) DMA, however, an infinite number of possibly prohibited differentiating attributes exists, which redoubles the complexity of the auditing procedure. On the one hand, the large number of possible discriminatory attributes makes it more difficult to find discrimination within the meaning of Article 6(5) DMA in the first place.[151] On the other hand, a larger number of protected attributes may also mean greater protection for the discriminated groups. Once a discriminatory attribute is identified, one could continuously expand the list of protected attributes by adding similar instances and intersections with other protected attributes. Thus, a higher and more differentiated level of protection could be achieved in the future. This, however, makes finding and remedying discrimination in the sense of the DMA computationally complex and costly.

### iv. Compliance requirements

Against this background, clear compliance guidelines must be offered. Article 8(1) DMA stipulates that gatekeepers do not only have the obligation to ensure compliance with the Regulation but should also be able to demonstrate this fact. Such compliance and documentation obligations are also prevalent in other highly regulated areas, such as banking law, the data protection law (Article 5(2) GDPR), and the proposed AI Act (e.g., Article 9 and 16 AI Act). However, particularly challenges arise for gatekeepers in the context of compliance with the fair ranking provision under Article 6(5) DMA.[152]

As seen, at least theoretically, an infinite amount of potentially discriminatory attributes exists, which would need to be considered. Fines, however, can only be imposed in cases of intent or negligence (Article 30(1) and (3) DMA). Hence, gatekeepers must have violated a duty of care to be liable. In our view, such a duty implies three distinctive compliance obligations.

---

[148] The fact that a company has entered into an exclusivity agreement with Amazon is not relevant to the accuracy of prediction, which is oriented towards the needs of consumers.
[149] Pessach/Shmueli, 'Algorithmic Fairness' (2020) arXiv preprint arXiv:2001.09784.
[150] Zehlike and others, 'FA*IR: A Fair Top-k Ranking Algorithm' in Proceedings of the 2017 ACM on Conference on Information and Knowledge, Management (Association for Computing Machinery 2017) 1569; Meike Zehlike, Philipp Hacker and Emil Wiedemann, 'Matching code and law: achieving algorithmic fairness with optimal transport' (2020) 34 Data Mining and Knowledge Discovery 163.
[151] Cf. Philipp Hacker, 'KI und DMA – Zugang, Transparenz und Fairness für KI-Modelle in der digitalen Wirtschaft' [2022] GRUR 1278, 1284.
[152] See also Cabral and others, 'The EU Digital Markets Act', (Joint Research Centre, JRC122910, Publications Office of the European Union 2021) 13.

The first one links Article 6(5) DMA even more closely to the transparency requirements in the P2B Regulation:[153] Gatekeepers need to constantly monitor the main parameters of the ranking, which also have to be disclosed. These features may not relate to any attributes protected under classical anti-discrimination law or to other illegal differentiating attributes (affiliation with the gatekeeper, etc.).

Second, gatekeepers must investigate evidence of possible discrimination in rankings brought to their attention, and remove the incriminated practice if warranted. This procedure essentially corresponds to the notice-and-takedown procedure applied in the context of potential copyright infringement in intellectual property law.[154] Its purpose lies in the establishment of a clearly structured and rapid procedure for checking and removing an unpredictable amount and type of potential legal violations.[155]

Intellectual property law, however, contains clear limitations of liability for hosting providers,[156] which the DMA lacks. This suggests, third, that gatekeepers are held to a higher level of care. Thus, in our view, the compliance requirements under the DMA also oblige gatekeepers to regularly examine their ML models for possible discrimination-relevant distinctions (test and audit), regardless of any specific allegations or clues that would suggest their existence.[157] This could, for example, be achieved by regularly creating two rankings for test purposes: one with and without the identification of the business users.[158] Should the two differ from each other, this could be construed as an indication of possible unjustified differentiations, as the ranking would therefore be more dependent on the mere identity of the customer, rather than on their attributes or those of their products.

Overall, the new ranking regulation in the DMA, while representing a step in the right direction, still contains significant challenges for effective implementation, both concerning regulatory agencies and addressees.

### III. Training data regulation

Whereas the previous section focused on the potential impact Article 6(5) DMA, and other pertaining regulation, might have on AI-based rankings, a number of other obligations in the DMA will also significantly affect the development and deployment of AI by gatekeepers and competitors. The first category in that context relates to restrictions regarding the use of certain data sets for gatekeepers' own purposes. This is relevant for AI training, validation and test data, which are a crucial component of AI modeling.[159]

---

[153] See also Brouwer (n. 78).
[154] See, e.g., Sharon Bar-Ziv and Niva Elkin-Koren, 'Behind the scenes of online copyright enforcement: Empirical evidence on notice & takedown' (2018) 50 Conn L Rev 339; Jeffrey Cobia, 'The digital millennium copyright act takedown notice procedure: Misuses, abuses, and shortcomings of the process' (2008) 10 Minn JL Sci & Tech 387.
[155] Jennifer M Urban and Laura Quilter, 'Efficient process or chilling effects-takedown notices under Section 512 of the Digital Millennium Copyright Act' (2005) 22 Santa Clara Computer & High Tech LJ 621, 622.
[156] See, e.g., Article 5 and 14 DSA and Article 17 C-DSM Directive.
[157] See also, for a practical proposal in this vein, Michael Veale and Reuben Binns, 'Fairer machine learning in the real world: Mitigating discrimination without collecting sensitive data' (2017) 4 Big Data & Society 2053951717743530.
[158] Cabral and others, 'The EU Digital Markets Act', (Joint Research Centre, JRC122910, Publications Office of the European Union 2021) 13.

### 1. Article 5(2) DMA

Article 5 (2) DMA contains a number of provisions relating to the cross-service use of personal end user data (PED). For example, gatekeepers may not use such data for personal advertising if it is obtained via third-party services that make use of gatekeepers' central platform services (lit. a). Furthermore, PED cannot be combined across services (lit. b), which specifically precludes the practice of combining data sets for enhanced training of AI systems. Even the registration of end users with other gatekeeper services by the gatekeepers themselves for the purpose of combining PED is prohibited (lit d). Finally, even without combining the data sets, PED may not be used by gatekeepers across services (lit. c). This would also include so-called federated learning strategies, in which the data remains formally separate, but the information obtained from them is combined in one single AI model.[160] Such strategies are, however, particularly privacy-preserving and sustainable in terms of energy consumption,[161] which is why their encumbrance should be reconsidered.

While these new rules have the potential to significantly affect gatekeepers' data management practices, the practical impact is considerably diminished by the exemption introduced in the same article, which allows these processing activities to occur if GDPR-compliant consent has been collected from the end user.[162] This invites legal uncertainty since numerous legal[163] and behavioral[164] problems related to the collection of valid and meaningful consent under the GDPR have been outlined in the literature. These mainly concern practices such as nudging,[165] bundling of purposes,[166] rational ignorance,[167] and "consent fatigue", which effectively causes individuals to simply agree to any given form of processing without even seeking to understand its implications.[168] The general inconsistency of users' statements regarding their privacy preferences and the actual online behavior is also referred to as the "privacy paradox".[169]

---

[159] Andreas Lindholm and others, Machine Learning - A First Course for Engineers and Scientists (CUP 2022) 67 ff, 299 et seqq.; Philipp Hacker, 'A legal framework for AI training data—from first principles to the Artificial Intelligence Act' (2021) 13 Law, Innovation and Technology 257, 259.
[160] Qiang Yang and others, 'Federated Machine Learning: Concept and Applications' (2019) 10 ACM Transactions on Intelligent Systems and Technology, Article 12.
[161] Basak Güler and Aylin Yener, 'Sustainable federated learning' (2021) arXiv preprint arXiv:210211274.
[162] See already Rupprecht Podszun, 'Should Gatekeepers Be Allowed to Combine Data? Ideas for Article 5(a) of the Draft Digital Markets Act' (2022) 71 GRUR Int 197, 199.
[163] See, e.g., Neil Richards and Woodrow Hartzog, 'The pathologies of digital consent' (2018) 96 Wash UL Rev 1461; Blume P, 'The inherent contradictions in data protection law' (2012) 2 International Data Privacy Law 26, 29 et seqq.
[164] Sheng Yin Soh, 'Privacy Nudges: An Alternative Regulatory Mechanism to "Informed Consen"' for Online Data Protection Behaviour' (2019) 5 EDPL 65; Yoan Hermstrüwer, 'Contracting around privacy: the (behavioral) law and economics of consent and big data' (2017) 8 J Intell Prop Info Tech & Elec Com L 9.
[165] C Utz and others, '(Un)informed Consent: Studying GDPR Consent Notices in the Field' in Proceedings of the 2019 ACM SIGSAC Conference on Computer and Communications Security (CCS '19) 973.
[166] Dominique Machuletz and Rainer Böhme, 'Multiple Purposes, Multiple Problems: A User Study of Consent Dialogs after GDPR' in Proceedings on Privacy Enhancing Technologies (2020) 481-498; Bojan Kostic and Emmanuel Vargas Penagos, 'The freely given consens and the "bundling" provision under the GDPR' (2017) 153 Computerrecht 217.
[167] Craig Van Slyke and others, 'Rational ignorance: A privacy pre-calculus' (2021) WISP 2021 Proceedings 12.

Considering how easily consent is often obtained in practice, the only significant restriction for gatekeepers therefore results from the modification of Article 6(1) GDPR, contained in Article 5(2)(3) DMA. It provides that where consent is refused, cross-service processing of personal data will only be possible where it can be based on compliance with a legal obligation, the protection of vital interests of a natural person or performance of a task in the public interest. This implies that the legal bases of performance of contract and legitimate interest (Article 6(1)(b) and (f) GDPR) can no longer be applied. Such a restriction is significant as these two legal bases are typically used as "fallbacks", where obtaining valid consent is impossible or challenging. Nevertheless, obtaining consent from rationally ignorant data subjects will often be quite feasible for gatekeepers.

For the moment, the overall impact of the rules seeking to prevent the accumulation of data by gatekeepers is, therefore, rather limited, especially since the gatekeepers will probably use the loopholes just discussed to their advantage.[170] While coupling the requirement for consent with the use of a service is not possible (Article 5(2)(2) DMA in conjunction with Article 7(4) GDPR[171]), businesses are still likely to present consent requests in a form that will ultimately lead a substantial amount of users to agree to cross-service processing.[172] The regulation does, however, look more promising when considering current developments concerning the prevention of 'dark patterns' seeking to steer users toward excessive consent.[173] The specification in Recital 67 DSA on dark patterns is aimed at a more neutral presentation of request, trying to assure that consent if in fact freely given and specific; ultimately, however, it defers to the UCPD in the GDPR.[174] Should this succeed, gatekeepers might be confronted with notable dropping in consent rates, potentially adding relevance to the DMA in the future. Importantly, the recent decision of the EDPB against Meta points exactly in this direction.[175] This adds to a general trend by Data Protection Authorities and NGOs to monitor and enforce the requirements for valid consent more aggressively.[176] To facilitate informed decisions by users, it would be worthwhile to combine this more robust enforcement with a traffic light system for different types of data sharing by gatekeepers.[177]

---

[168] Nouwens and others, 'Dark Patterns after the GDPR: Scraping Consent Pop-ups and Demonstrating their Influence' in Proceedings of the 2020 CHI Conference on Human Factors in Computing Systems (2020) 1.
[169] S Barth and MDT de Jong, 'The privacy paradox – Investigating discrepancies between expressed privacy concerns and actual online behavior – A systematic literature review' (2017) 34 Telematics and Informatics 1038 – 1058.
[170] D Geradin, K Bania and T Karanikioti, 'The interplay between the Digital Markets Act and the General Data Protection Regulation' (August 29, 2022) <http://dx.doi.org/10.2139/ssrn.4203907> accessed: 06 November 2022.
[171] Bojana Kostic and Emmanuel Vargas Penagos, 'The freely given consent and the "bundling" provision under the GDPR' (2017) 153 Computerrecht 217; Frederik Zuiderveen Borgesius and others, 'Tracking walls, take-it-or-leave-it choices, the GDPR, and the ePrivacy regulation' (2017) 3 Eur Data Prot L Rev 353, 361.
[172] Cf. C Utz and others, '(Un)informed Consent: Studying GDPR Consent Notices in the Field' in Proceedings of the 2019 ACM SIGSAC Conference on Computer and Communications Security (2019) 973.
[173] See, e.g., Jamie Luguri and Lior Strahilevitz, 'Shining a light on dark patterns' (2021) 13 Journal of Legal Analysis 43; Mario Martini et al., 'Dark patterns' (2021) 1 Zeitschrift für Digitalisierung und Recht 47.
[174] See the analysis in Philipp Hacker, 'Manipulation by algorithms. Exploring the triangle of unfair commercial practice, data protection, and privacy law' (2022) European Law Journal.
[175] noyb, 'noyb win: Personalized Ads on Facebook, Instagram and WhatsApp declared illegal' (noyb 6 December 2022) <https://noyb.eu/en/noyb-win-personalized-ads-facebook-instagram-and-whatsapp-declared-illegal> accessed 8 December 2022.
[176] Jennifer Bryant, "Belgian DPA fines IAB Europe 250K euros over consent framework GDPR violations' (iapp 2 February 2022) <https://iapp.org/news/a/belgian-dpa-fines-iab-europe-250k-euros-over-consent-framework-

### 2. Article 6(2) DMA

A corresponding prohibition, aimed at business user data, can be found in Article 6(2) DMA: gatekeepers are prohibited from using data in competition with business users that these or their end users have generated or provided in the context of relevant core platform services. This rule, however, does not apply to data that is already publicly accessible. To delineate that concept, Article 6(2)(2) DMA specifies that non-public data also includes information that can be "inferred from, or collected through, the commercial activities of business users or their customers". Consequently, it will neither be possible to use end user or business user data for AI-based inferences, nor to further harness them in competition with business users.

The introduction of the new restriction makes sense from the perspective of workable competition as it precludes a further entrenchment of the gatekeeping position via data-based inferences. The EU, with this rule, moves onto largely uncharted territory since lawmakers have, until this point, largely refrained from regulating AI-based inferences, in spite of their economic and informational importance[178] and criticism from literature.[179]

The overall impact of Article 6(2) DMA will probably be stronger than that of Article 5(2) DMA as the former applies to all types of data, not just personal data. Given the practical difficulties in distinguishing personal from non-personal data, especially in the context of AI training data,[180] this simplification is a step in the right direction. Additionally, it should be noted that Article 6(2) DMA, in contrast to Article 5(2) DMA, cannot be waived based on consent or any other grounds, expanding its impact even further.

### 3. Comparison with Article 10 AI Act

Contrasting the DMA data governance regime with Article 10 AI Act, the provision detailing specific requirements for data used in training high-risk AI applications,[181] reveals strikingly

---

gdpr-violations/> accessed 2 September 2022; NOYB, 226 Complaint Lodged Against Deceptive Cookie Banners, <https://noyb.eu/en/226-complaints-lodged-against-deceptive-cookie-banners> accessed: 21 September 2022.
[177] See also Rupprecht Podszun, 'Should Gatekeepers Be Allowed to Combine Data? Ideas for Article 5(a) of the Draft Digital Markets Act' (2022) 71 GRUR Int 197, 201 et seq.; Philipp Hacker, Datenprivatrecht (Mohr Siebeck 2020), 627 et seqq. on privacy scores.
[178] See, e.g., Avi Goldfarb, Shane Greenstein and Catherine Tucker, 'Introduction to Economic Analysis of the Digital Economy, in id. (eds), Economic Analysis of the Digital Economy (University of Chicago Press 2015), 1.
[179] See only Sandra Wachter and Brent Mittelstadt, 'A right to reasonable inferences: re-thinking data protection law in the age of big data and AI' (2019) 2 Colum Bus L Rev 494.
[180] See, e.g., Philipp Hacker, 'A legal framework for AI training data—from first principles to the Artificial Intelligence Act' (2021) 13 Law, Innovation and Technology 257, 265-268; see also, more generally, Michèle Finck and Frank Pallas, 'They who must not be identified—distinguishing personal from non-personal data under the GDPR' (2020) 10 International Data Privacy Law 11.
[181] For a critical analysis of Article 10 AI Act, see Philipp Hacker, 'A legal framework for AI training data—from first principles to the Artificial Intelligence Act' (2021) 13 Law, Innovation and Technology 257, 296-300; Marvin van Bekkum and Frederik Zuiderveen Borgesius, 'Using sensitive data to prevent discrimination by artificial

different objectives. Article 10 AI Act compels developers of high-risk AI systems to only use high-quality training data to facilitate the creation of accurate predictions and to mitigate bias in AI systems. For that purpose, several quality criteria are outlined in Article 10(2) to (5) AI Act, inter alia specifying the representativeness of training data for the target group and statistical appropriateness. In contrast, the DMA restrictions seeks to prevent gatekeepers from gathering high-quality data sets by tapping into the data trove accumulating on the platform; if anything, this will reduce predictive accuracy of gatekeeper models.

The DMA is none of the less right in explicitly blocking gatekeepers from leveraging their specific position to build better models. This would most likely have the effect of even further cementing their position on the market, thereby continuing to hinder the possibility of workable competition. This points to an inherent conflict of objectives in the area of AI and platform regulation: technological tools and AI systems are supposed to be high-performing, but it is precisely this capability that may lead to further market concentration and the weakening of competitive processes. The AI Act and the DMA, therefore, rightly seek to accommodate this tension by allocating specific and, prima facie, strikingly divergent duties to gatekeepers on the one hand and developers of high-risk AI systems on the other.

### IV. Access rights

Another set of AI-relevant rules targeting the competitive position of gatekeepers are the access rights contained in Article 6(10) and 11 DMA.[182] The idea of harnessing access rules, instead of a data producers' right, to strengthen innovation and competition in data-driven markets has already received much scholarly attention.[183] The DMA has now, for the first time, introduced general access rules for gatekeepers, independently of the business sector in which they are active.

In a sense, access rights are the flip side of the restrictions concerning the use of training data. While these limitations are supposed to prevent gatekeepers' AI systems from becoming too powerful, creating an unfair advantage, access rights are intended to provide business users with the necessary tools to develop high-performing algorithms themselves, including machine learning models. Since access rights are not considered in the current draft of the AI Act, the importance of such within the DMA is even more pronounced.

#### 1. Article 6(10) DMA

According to Article 6(10) DMA, business users or authorized third parties may have access to data provided for or generated in the context of the use of the respective core platform services by the business users themselves or their end users. Access must be granted free of charge and

---

intelligence: Does the GDPR need a new exception?' (2023) 48 Computer Law & Security Review 105770, 11-12.
[182] The right of access in Article 6 para. 12 DMA, on the other hand, has no specific reference to AI.
[183] Daniel L. Rubinfeld, Michal S. Gal, 'Access Barriers to Big Data' (2017) 59 Ariz. L. Rev. 339; Wolfgang Kerber, 'Governance of Data: Exclusive Property vs. Access' (2016) 47 IIC-International Review of Intellectual Property and Competition Law 759; see also Schweitzer, 'Datenzugang in der Datenökonomie: Eckpfeiler einer neuen Informationsordnung' (2019) 121 Gewerblicher Rechtsschutz und Urheberrecht, 569

in a way that is effective, high-quality, continuous and real-time. In that context, AI could be particularly helpful in forecasting demand and optimizing product design.

The provision does, however, also include important restrictions regarding access to personal data. Given the broad interpretation of the concept of personal data by the CJEU and legal scholarship,[184] this concerns a large share of the data eligible to be accessed. First of all, it may only be provided where the information is directly related to the use of the business user's services and products by the end user, through the relevant core platform service. This, in turn, means that businesses will still only receive information concerning individuals who are already part of their client base. In that sense, the competitive effect will be significantly limited: information about consumers the business user was not yet able to reach would arguably be more valuable in that respect.

Second, gatekeepers are only allowed to share such personal data where the user has provided their consent. As has already been mentioned, businesses have by now found a number of ways to assure the obtainment of consent from data subjects through the use of means such as nudging, specific framings or bundling of purposes.[185] If gatekeepers design and steer user consent, they should be able to guide them into excluding the data use by business users. Such behavior would arguably be legal, but render Article 6(10) DMA effectively futile, as all data generated by a customer will fall under the definition of personal data and, hence, the outlined restriction. This issue will be taken up again in the final part of the paper (VI.2.).

## 2. Article 6(11) DMA

The provisions introduced by Article 6(11) DMA, by contrast, intend to remedy a central weakness of the market for search engine operators. It is common knowledge that the quality of a search engine is predominantly determined based on the delivered results. These are however, largely being optimized by means of analyzing the historical search and click behavior of end customers, with the support of machine learning systems.[186] Therefore, whichever provider is able to generate more end users up front will naturally be able to increase their competitive advantage even further, potentially creating a positive feedback loop of extended competitive advantage on the side of the gatekeeper.

---

[184] See, e.g., Michèle Finck and Frank Pallas, 'They who must not be identified—distinguishing personal from non-personal data under the GDPR' (2020) 10 International Data Privacy Law 11; Nadezhda Purtova, 'The law of everything. Broad concept of personal data and future of EU data protection law' (2018) 10 Law, Innovation and Technology 40.
[185] Dominique Machuletz, Rainer Böhme, 'Multiple Purposes, Multiple Problems: A User Study of Consent Dialogs after GDPR' Proceedings on Privacy Enhancing Technologies 2020, 481-498, http://dx.doi.org/10.2478/popets-2020-0037; Bojana Kostic, Emmanuel Vargas Penagos, 'The freely given consens and the 'bundling' provision under the GDPR' (2017) 153 Computerrecht, 217; Christine Utz et al., '(Un)informed Consent: Studying GDPR Consent Notices in the Field', Proceedings of the 2019 ACM SIGSAC Conference on Computer and Communications Security (CCS '19), 973–990, https://doi.org/10.1145/3319535.3354212; Midas Nouwens et al., 'Dark Patterns after the GDPR: Scraping Consent Pop-ups and Demonstrating their Influence', Proceedings of the 2020 CHI Conference on Human Factors in Computing Systems (CHI '20), 1–13, https://doi.org/10.1145/3313831.3376321.
[186] Daniel L. Rubinfeld, Michal S. Gal, 'Access Barriers to Big Data' (2017) 59 Ariz. L. Rev. 339, 353; Martens, 'An Economic Policy Perspective on Online Platforms' (2016) Institute for Prospective Technological Studies Digital Economy Working Paper 2016/05, 4, 24 et seq.; see also n. 19.

To mitigate the impact of this phenomenon, Article 6(11) DMA conveys upon search engine operators the right to access the data set of gatekeepers who themselves operate search engines. The access to these ranking, query, click and view data must be provided on FRAND terms, which have already been outlined above, the purpose again being the enablement of business users to optimize their own AI models (Recital 61 DMA).

Significantly, Article 6(11) DMA also provides that any personal data part of the ranking, query, click and view data must be offered in an anonymized form, meaning that it cannot be related to an identified or identifiable natural person.[187] While the introduction of such an obligation is definitely reasonable from the perspective of protecting the individuals' rights and freedoms, it does create significant implementation issues on the side of the gatekeeper. The anonymization of personal data is, after all, not an uncomplicated task, as research has repeatedly shown that supposedly anonymized data can be de-anonymized through a range of strategies.[188]

Consequently, it will be necessary for gatekeepers to use strong, state-of-the-art anonymization strategies to comply with the requirements of the DMA and the GDPR. Since, however, strong anonymization also requires a fair amount of data to be either removed from or modified in the existing set,[189] such measures can also lead to a reduction in value of the data set for the purpose of AI training.[190] In that sense so-called privacy-preserving machine learning (PPML) strategies,[191] which attempt to strike a balance between data protection and performance, are likely to gain more practical and regulatory relevance as a result of Article 6(11) DMA. On the other hand, it is questionable to what extent gatekeepers would be compelled to go above and beyond in the search for privacy preserving techniques, assuring a high-quality data set for their competitors. At least in the initial phase, businesses are likely to simply provide extensively altered and anonymized data sets, with limited practical use. As PPML progresses, so does the requirement for gatekeepers to offer data sets that are not only anonymized but also performance-enabling due to state-of-the-art PPML.

### 3. Comparison with Data Act Proposal

Another legal basis introducing extensive access rights on the EU level is the Data Act (DA),[192] which serves the purpose of removing barriers in data sharing.[193] The access rights under the

---

[187] Michèle Finck and Frank Pallas, 'They who must not be identified—distinguishing personal from non-personal data under the GDPR' (2020) 10 International Data Privacy Law 11, 15.
[188] See, e.g., Luc Rocher, Julien M. Hendrickx, Yves-Alexandre de Montjoye, 'Estimating the success of re-identifications in incomplete datasets using generative models' (2019) 10 Nature Communications, Art. 3069.
[189] See, e.g., Irish Data Protection Commission, 'Guidance on Anonymisation and Pseudonymisation' (June 2019), <https://www.dataprotection.ie/sites/default/files/uploads/2019-06/190614%20Anonymisation%20and%20Pseudonymisation.pdf> accessed 7 December 2022; Article 29 Data Protection Working Party, Opinion 05/2014 on Anonymisation Techniques, WP 216, 2014.
[190] Villaronga et al., 'Humans forget, machines remember: Artificial intelligence and the Right to Be Forgotten' (2018) 34 Computer Law & Security Review 304, 310.
[191] See, e.g., Payman Mohassel and Yupeng Zhang,'SecureML: A System for Scalable Privacy-Preserving Machine Learning' (2017) IEEE Symposium on Security and Privacy (SP) 2017, 1.
[192] European Commission, Proposal for a Regulation on the European Parliament and of the Council on harmonised rules on fair access to and use of data (Data Act) COM(2022) 68 final.
[193] Data Act (n. 192), 2 et seq.

Data Act are, however, significantly more consumer-driven than those under the DMA.[194] Articles 4 and 5 DA regulate the right of users to access and use data generated by the use of products or related services and to share it with third parties, respectively. Other businesses will therefore potentially be given the option of using such information, but only "upon request by a user, or by a party acting on behalf of a user".[195] Hence, the consumer would need to act first, or at least clearly communicate their preferences. The DA seeks to balance the acquisition of data by competing businesses with individual's right to self-determination.[196] The only instances in which a right to access is directly recognized for a legal entity are regulated in Chapter V and benefit public sector bodies or Union institutions, agencies or bodies, for example in cases of emergencies.[197]

This dependency on the consumer's initiative, while conducive to the individual's rights and freedoms, also makes the access rights under the DA significantly less useful when it comes to providing business users with the necessary tools to develop powerful analytical methods themselves, including machine learning models.[198] After all, the value of the data sets kept by gatekeepers lies largely in their comprehensiveness. If business users are dependent on first agreeing on the conditions of the processing with the end user (Article 6(1) DA), the collection of a data set, with a wide enough scope to be relevant, will be considerably more difficult due to simple transaction costs.

Significantly, though, the aims pursued by the DMA are also recognized by the DA.[199] First, directly through Article 5(2) DA, which excludes businesses designated as gatekeepers from the possibility of making use of the outlined access rights.[200] Second, indirectly through Article 7(1) DA, which provides that micro and small enterprises do not need to accommodate such access. This clearly underlines the overarching purpose of both acts to even out current imbalances on the market. The extent to which they will achieve their common goal, however, remains doubtful, as the last section of paper explores further (V.2.).

## V. Information requirements regarding advertising

Last but not least, Articles 5 and 6 DMA also contain several information requirements for gatekeepers regarding their advertising practices. Their primary purpose is to counteract the information asymmetry between platforms and business users concerning the conditions and functioning of advertisements.[201] While the requirements themselves are arguably the least intrusive of those concerning data and AI, they are still crucial for many business users. Under the current market realities, advertising constitutes the central source of revenue for most platforms, including gatekeepers, as well as their competitors.[202] The Commission is currently considering tightening the EU acquis even further with a view to fair advertising.[203] Given their economic significance, the potential impact of the DMA rules on advertising will be examined in the following sections, and compared to the information requirements in the AI Act.

---

[194] Regarding the relationship of the DA to other legal acts including DMA cf. also Louisa Specht-Riemenschneider, 'Der Entwurf des Data Act - Eine Analyse der vorgesehenen Datenzugangsansprüche im Verhältnis B2B, B2C und B2G' (2022) 25 Zeitschrift für IT-Recht und Recht der Digitalisierung 809, 810 et seq; critical with regard to the achievement of consumer empowerment Wolfgang Kerber, 'Governance of IoT Data: Why the EU Data Act Will not Fulfill Its Objectives' (2022) GRUR International 1.
[195] See Art. 5(1) DA.
[196] See, for a critique regarding the effectiveness of the DA's guarantee of self-determination, Louisa Specht-Riemenschneider (n. 194), 816 et seqq.
[197] On the B2G ("business to government") relationship see also Louisa Specht-Riemenschneider (n. 194) 824 et seqq.
[198] For further critique on the dependency on the consumer's initiative see, e.g., Rupprecht Podszun and Philipp Offergeld, The EU Data Act and the Access to Secondary Markets (October 24, 2022). Available at SSRN: https://ssrn.com/abstract=4256882, 45 et seq.
[199] Data Act (n. 192), 5.
[200] See also Inge Graef and Martin Husovec, 'Seven Things to Improve in the Data Act' (2022) SSRN: https://ssrn.com/abstract=4051793, 2 et seq., who mention the possibility of bypassing the exclusion by relying on Article 20 GDPR instead.

### 1. Article 5(9) and (10) DMA

Article 5(9) and (10) DMA oblige gatekeepers to disclose to both advertisers and publisher, respectively, upon request, the metrics used to calculate prices, fees, and remunerations for each advertisement placed or displayed. Metrics used in the field of advertisement, and e-commerce in particular, are both numerous and varied.[204] Also, based on their crucial influence on the business model, enterprises are constantly working on their optimization.[205] The impact of a disclosure requirement on gatekeepers should not be underestimated, especially when taking into consideration the fact that prices for advertising are typically determined automatically within fractions of a second by real-time auctions, which in turn are backed by complex algorithms or machine learning techniques.[206] Concerning AI, this new obligation therefore indirectly compels gatekeepers, and ad tech networks,[207] to use explainable AI Systems.[208]

The rule also has another, quite significant impact. Since the requirement refers to each individual advertisement, this effectively creates an obligation to deliver local explanations, which disclose the relevant features for each individual decision.[209] While this can be quite burdensome, it is increasingly possible even with complex, 'black box' systems such as artificial neural networks, as seen.[210] A similar obligation might have been considered potentially too burdensome for regular companies.[211] Its fulfilment by gatekeepers should, however, be considered a proportionate measure.

---

[201] See, for example, Damien Geradin and Dimitrios Katsifis, 'An EU competition law analysis of online display advertising in the programmatic age' (2019)15 European Competition Journal 55, 62; Recital 45 DMA.
[202] Damien Geradin and Dimitrios Katsifis, 'An EU competition law analysis of online display advertising in the programmatic age' (2019)15 European Competition Journal 55, 55 f.
[203] Luca Bertuzzi, 'Dark patterns, online ads will be potential targets for the next Commission, Reynders says' EURACTIV (December 9, 2022), <https://www.euractiv.com/section/digital/interview/dark-patterns-online-ads-will-be-potential-targets-for-the-next-commission-reynders-says> accessed December 9, 2022.
[204] Xiangyu Liu et al., 'Neural Auction: End-to-End Learning of Auction Mechanisms for E-Commerce Advertising' (2021) Proceedings of the 27th ACM SIGKDD Conference on Knowledge Discovery and Data Mining, 3354-3364 <https://doi.org/10.1145/3447548.3467103> 9 December 2022; Zhilin Zhang et al., 'Optimizing Multiple Performance Metrics with Deep GSP Auctions for E-commerce Advertising' (2021) Proceedings of the Fourteenth ACM International Conference on Web Search and Data Mining (WSDM '21), 993-1001 <https://doi.org/10.1145/3437963.3441771> 9 December 2022; Sahin Cem Geyik et al., 'Joint Optimization of Multiple Performance Metrics in Online Video Advertising' (2016) Proceedings of the 22nd ACM SIGKDD International Conference on Knowledge Discovery and Data Mining (KDD '16), 471–480. <https://doi.org/10.1145/2939672.2939724> 9 December 2022.
[205] Cf. ibid.
[206] Damien Geradin and Dimitrios Katsifis, 'An EU competition law analysis of online display advertising in the programmatic age' (2019)15 European Competition Journal 55, 61.
[207] See, on their significance, see Damien Geradin D, 'GDPR Myopia: how a well-intended regulation ended up favouring large online platforms-the case of ad tech' (2021) 17 European Competition Journal 47, 50 et seqq.
[208] See the overview in n. 79.
[209] See n. 47 and accompanying text.
[210] See n. 46 and accompanying text.

### 2. Article 6(8) DMA

Similarly, Article 6(8) DMA requires gatekeepers to disclose, to advertisers and publishers, the tools they use to measure the performance of advertising. In instances where AI systems are being deployed, these will often be performance metrics, such as predictive accuracy.[212] Additionally, the data necessary for business users to perform their own verifications must be made available. In the context of advertising, this will include data such as the conversion rate (*click-through rate*).[213]

### 3. Alignment with the AIA (Article 11, 13, 15)

In comparison with the transparency requirements in Articles 13 and 11 AI Act in conjunction with Annex IV AI Act, it is noteworthy that the information owed according to the DMA has a significantly narrower scope of application, as it only concerns advertising. Conversely, the type of information which needs to be provided is more detailed, as businesses will be entitled to receive local explanations, concerning each individual case.[214] In the DMA, only global explanations, concerning the entire model, are foreseen (see above, Part II.2.f)).

However, a right to a more "concrete" explanation will not necessarily benefit gatekeepers' competitors, who might have been more interested in an overall explanation of the functioning of the underlying AI system. On the other hand, the DMA provides businesses with a completely new claim, as a right to a local explanation is either not foreseen or heavily debated under other EU acts (see above, Part II.2.). While the option to receive information on individual advertisements provides advertisers and publishers with more choice, it is, from a systematic perspective, surprising to see this included in a piece of legislation aimed at remediating distortions in competition, as local approaches are generally preferred by users, whereas business users and developers should have a greater interest in global explanations.[215] Furthermore, it should be acknowledged that the verification efforts enabled by the DMA partly presuppose considerable technical prowess on the part of advertisers and publishers, which will not always be present to a sufficient extent.

Nevertheless, the overall aim of the DMA to ensure greater transparency in the advertising market and in AI systems used in it should be welcomed, both from a competition and a due process perspective.[216] In its current form, it has potential to facilitate the comparison of advertising conditions and the verification of a platform's promises, fostering the contestability of gatekeeper positions in the advertising and publishing market.

---

[211] Cf. Damien Geradin D, 'GDPR Myopia: how a well-intended regulation ended up favoring large online platforms-the case of ad tech' (2021) 17 European Competition Journal 47, 48-49.
[212] X.Liu et al. (n. 204); Z.Zhang et al. (n. 204).
[213] See also n. **Fehler! Textmarke nicht definiert.** and accompanying text.
[214] Cf. for example Annex IV para. 2 lit. b AI Act: meaning of the various parameters; see also Philipp Hacker and Jan-Hendrik Passoth, 'Varieties of AI Explanations Under the Law. From the GDPR to the AIA, and Beyond' in Holzinger et al. (eds), xxAI – Beyond Explainable AI, International Workshop on Extending Explainable AI Beyond Deep Models and Classifiers (Springer 2022), 343, 357 et seqq.
[215] Cf. The Royal Society, Explainable AI: The Basics – Policy Briefing, November 2019, <https://royalsociety.org/topics-policy/projects/explainable-ai/> accessed 2 October 2022, 14.
[216] Cf. also Danielle Citron and Frank Pasquale, 'The scored society: Due process for automated predictions' (2014) 89 Wash L Rev 1; Danielle Citron, 'Technological due process' (2007) 85 Wash UL Rev 1249.

**VI. Regulating gatekeeper data and AI going forward**

The preceding sections have revealed that the DMA complements a growing and increasingly elaborate regime regulating data, algorithms, and models used by gatekeepers in performing their core business functions. However, the analysis has also shown that significant shortcomings remain. The final part of the paper therefore sketches amendments and policy proposals regarding the three areas covered in this essay: transparency, access, and fairness.

### 1. Transparency

With respect to transparency, the DMA adds to an already copious, but incoherent regime demanding various types of transparency activities from platforms and AI providers or users. The AI Act is bound to complicate this regime with further disclosure and explainability duties.

What is lacking, so far, is a unifying framework for transparency and explainability with respect to complex software systems, including AI. This gap is detrimental for both end users and AI developers as it entails legal uncertainty, raises the cost of compliance and litigation, and fails to meet the purpose of the transparency regime: balancing the fundamental right of data protection and access to information, both of consumers and business users, with countervailing rights and interests of gatekeepers.

#### a) Relationship to XAI

With the DMA now enacted, the AI Act would have the unique opportunity to consolidate the EU algorithmic transparency regime, specifically for AI and complex software falling under the broad AI definition of Article 3(1) AI Act, read in conjunction with Recitals 6a and 6b AI Act.[217] To foster innovation and legal certainty–whose absence is quite detrimental to AI development and deployment–the requirements for explanations should be further specified and adapted to varying recipients.

##### i. Opening the black box

As mentioned, there are many techniques, developed over the last years in computer science research, for opening so-called "black box" AI systems long thought to be particularly opaque.[218] However, not all of these techniques fit the needs of all audiences; rather, they must be actionable.[219] End users, for example, will primarily be interested in ascertaining that the rankings created by gatekeepers reflect their interests and preferences, and not those of the gatekeeper. To this end, feature salience can indeed be helpful. If the main features list

---

[217] See n. 1.
[218] For an overview, see references in n. 79.
[219] Philipp Hacker and Jan-Hendrik Passoth, 'Varieties of AI Explanations Under the Law. From the GDPR to the AIA, and Beyond' in Holzinger et al. (eds), xxAI – Beyond Explainable AI, International Workshop on Extending Explainable AI Beyond Deep Models and Classifiers (Springer 2022), 343, 362 et seqq.

"affiliation with gatekeeper" or other categories unrelated to consumer preference, end users may switch to other providers.

Even though many customers will probably ignore such disclosures,[220] they will nevertheless be analyzed by consumer associations, journalists, or even regulatory agencies who may then act as information intermediaries, or enforcement entities, for any suboptimal or illegal features found.[221] Increasingly, such analysis is automated using machine learning as well, so that the amount of data analyzed and the policies flagged as problematic increase substantially.[222] This, in turn, increases deterrence and compliance pressure.

### ii. Actionable explanations for business users

For business users, in turn, feature salience is important as well: they may deduce if their products were ranked in a meaningful and fair way. Business users also have a much greater incentive than consumers to monitor ranking conditions as these are essential for commercial success on the platform. Opaque rankings or main features unrelated to product performance will raise incentives to switch platforms, fostering the contestability of rankings markets.

Moreover, however, business users will also be interested in how they may improve the ranking of their products. Technically, this may be achieved by delivering so-called counterfactual explanations.[223] In this situation, however, they raise the question of possible manipulations of the ranking, to which we now turn.

### b) Trade Secrets and Manipulation

From a legal point of view, the mentioned disclosure requirements also raise the problem of trade secrets and the manipulability of the ranking. This needs to be considered in transparency rules concerning AI and software more generally.

### i. Platform problems

First, far-reaching transparency rules may undermine incentives to invest in innovation in the first place, and counteract the protection afforded by IP rights. Competitors may reverse engineer algorithms, models or data sets and free-ride on discoveries made by the holders of trade secrets.[224] However, the empirical evidence of the fact of trade secrets on innovation is mixed. While some results suggest that trade secrets spur investment in research and development, particularly in the IT sector,[225] another recent empirical study suggests that strengthened trade secret regimes may in fact hamper, and not foster, innovation.[226] In markets dominated by informal networks of learning and collaboration, strong trade secrets protection often does more harm than good by preventing the exchange of ideas and knowledge.[227] AI research and development are, arguably, to a considerable extent based on such networks. Hence, claims to protect trade secrets for the sake of AI development should be taken *cum grano salis*. In the specific situation of business users seeking information about data and models used by gatekeepers, the argument of protecting trade secrets is weakened even further as the market would generally benefit from greater contestability and competition, challenging the entrenched position of gatekeepers.

---

[220] Cf. Omri Ben-Shahar and Adam Chilton, 'Simplification of privacy disclosures: an experimental test' (2016) 45 The Journal of Legal Studies S41; Jonathan A. Obar and Anne Oeldorf-Hirsch, 'The biggest lie on the internet: Ignoring the privacy policies and terms of service policies of social networking services' (2020) 23 Information, Communication & Society 128.
[221] See, e.g., Paul Bischhoff, 'Comparing the privacy policy of internet giants side-by-side (comparitech March 2017) <https://www.comparitech.com/blog/vpn-privacy/we-compared-the-privacy-policies-of-internet-giants-side-by-side/> accessed 6 September 2022; Forbrukerrådet, Deceived by Design, Report, 2018, <https://fil.forbrukerradet.no/wp-content/up/ loads/2018/06/2018-06-27-deceived-by-design-final.pdf>, accessed 6 September 2022.
[222] See, e.g., Sebastian Zimmeck and Steven M. Bellovin, 'Privee: An architecture for automatically analyzing web privacy policies', (2014) 23rd USENIX Security Symposium 1; Lisa Austin et al., Towards Dynamic Transparency: The AppTrans (Transparency for Android Applications) Project, Working Paper, 2018, https://ssrn.com/abstract=3203601.
[223] Sandra Wachter, Brent Mittelstadt and Chris Russell, 'Counterfactual explanations without opening the black box: Automated decisions and the GDPR' (2017) 31 Harv JL & Tech 841; Ramaravind Kommiya Mothilal, Amit Sharma, Chenhao Tan, 'Explaining machine learning classifiers through diverse counterfactual explanations' (2020) Proceedings of the 2020 conference on Fairness, Accountability, and Transparency 607.

Second, providing information about the main factors relevant for ranking provides opportunities for their manipulation.[228] This problem is particularly virulent if the predictive features are merely correlated with and not causal for the target variable. In this case, they can be artificially changed by business users and cause an improvement in the ranking position without any concurring improvement in the characteristics the target variable seeks to capture. For example, in one study, the purchase of felt tips (for preventing damage to the floor by moving furniture) was found to be highly predictive for creditworthiness. The relationship is obviously a mere correlation, not a causal one. If it was disclosed that a credit scoring model takes this into account, candidates could order felt tips on purpose to improve their creditworthiness.

Ultimately, however, this does not speak against transparency per se, but in favor of using causal inference instead of models based primarily on correlations. Nevertheless, while causal machine learning is making steady progress,[229] it cannot be deployed across the board yet.[230] As long as correlational models persist as the state-of-the-art technology in many areas, manipulability needs to be taken into account when designing transparency rules.

### ii. Legal solutions

Article 5(6) of the P2B Regulation explicitly considers trade secrets and manipulability. It exempts the disclosure of such information from the general transparency obligation that makes it possible to manipulate the ranking with sufficient probability, and mentions the Trade Secrets

---

[224] David D. Friedman, William M. Landes and Richard A. Posner, 'Some economics of trade secret law' (1991) 5 Journal of Economic Perspectives 61.
[225] Ivan Png, 'Law and innovation: evidence from state trade secrets laws' (2017) 99 Review of Economics and Statistics 167.
[226] Andrea Contigiani, David H. Hsu and Iwan Barankay, 'Trade secrets and innovation: Evidence from the "inevitable disclosure" doctrine' (2018) 39 Strategic Management Journal 2921.
[227] Laura Pedraza-Fariña, 'Spill your (trade) secrets: Knowledge networks as innovation drivers' (2016) 92 Notre Dame L Rev 1561.
[228] Jane Bambauer and Tal Zarsky, 'The Algorithm Game' (2018) 94 Notre Dame Law Review 1.
[229] See, e.g., Judea Pearl, 'The seven tools of causal inference, with reflections on machine learning' (2019) 62 Communications of the ACM 54; Jonathan Richens et al., 'Improving the accuracy of medical diagnosis with causal machine learning' (2020) 11 Nature Communications 1.
[230] Jean Kaddour et al., 'Causal machine learning: A survey and open problems' (2022) arXiv preprint arXiv:220615475.

Directive.[231] Still, it does not provide guidelines for balancing the need for transparency with these countervailing interests. What needs to be disclosed are the features and their relative importance. The described tension can be resolved, in our view, in such a way that no precisely quantified weights are divulged, but only intervals or even only ordinally ordered lists of the relevance of the individual parameters.[232] In this way, it will be distinctively more difficult to reverse engineer the model, and the risk of manipulability is also lowered.

Importantly, one will have to apply the manipulation and trade secret protection of Article 6(5) P2B Regulation to Article 7 of the P2B Regulation by analogy, since manipulation and free-riding can occur just as much based on the information provided in the T&C.

Turning to the GDPR, Article 15 GDPR, as all fundamental and GDPR rights, does not apply without restrictions. Next to the general exemptions of manifestly unfounded and excessive requests, contained in Article 12(5) GDPR, Article 15(4) GDPR also postulates that rights and freedoms of others should not be adversely affected. While the clause technically only refers to the right to obtain a copy (Article 15(3) GDPR), it should by analogy and where necessary apply to the details provided under Article 15(1) GDPR as well. This understanding is also supported by Sentence 5 of Recital 63 GDPR and the Guidelines of the Article 29 Working Party.[233]

This analogy is of particular relevance to the scope of access to the underlying model. While rights of others in the sense of different data subjects are unlikely to be affected, the trade secrets or intellectual property rights of the gatekeeper might in fact be engaged and warrant a restriction of the individual's rights.[234] Significantly, this does not imply that, in such instances, no information should be provided. Rather, the applicable rights need to be balanced. For example, less intrusive means of access, such as partial access or less granular information, should be considered, like in the case of information according to the P2B Regulation.

### 2. Access

For consumers, transparency, for example, an explanation of a decision, is an important prerequisite to contesting data practices. Business users, however, additionally need access to data, and potentially models, to use them for products that may eventually challenge the gatekeepers' competitive position. The access rights contained in the DMA, and the DA, are steps in the right direction. However, they still do not go far enough.

First, as seen, access to end user data under Article 6(10) DMA hinges on end user consent, which is generally collected by gatekeepers. They have, however, an incentive to discourage consent in this respect in respect in order to block access requests by potential competitors.

---

[231] See also Christoph Busch, 'Mehr Fairness und Transparenz in der Plattformökonomie? Die neue P2B-Verordnung im Überblick' (2019) 121 Gewerblicher Rechtsschutz und Urheberrecht 788, 793.
[232] See also Bäcker, in: Kühling/Buchner, DS-GVO, 2020, 3rd ed., Article 13 para. 54; Philipp Hacker and Jan-Hendrik Passoth, 'Varieties of AI Explanations Under the Law. From the GDPR to the AIA, and Beyond' in Holzinger et al. (eds), xxAI – Beyond Explainable AI, International Workshop on Extending Explainable AI Beyond Deep Models and Classifiers (2022), 343, 350.
[233] Article 29 Data Protection Working Party, 'Guidelines on Automated individual decision-making and Profiling for the purposes of Regulation 2016/679' (2017) WP 251, 17.
[234] Cf. also Article 29 Data Protection Working Party, 'Guidelines on Automated individual decision-making and Profiling for the purposes of Regulation 2016/679' (2017) WP 251, 17.

Hence, business users should be allowed to review and veto the design of consent requests by gatekeepers in so far as consent to the reuse of personal data by their own end users according to Article 6(10) DMA is concerned. In such a setting, business users would be able to reject consent designs under which gatekeepers, via framing or other behavioral effects,[235] seek to obtain consent for their own data sharing practices but to nudge users to withhold consent regarding data sharing with business users.

Furthermore, an amendment to Article 6(11) DMA should specify that personal data cannot be anonymized arbitrarily by gatekeepers, but only in a way that preserves the utility of the data set to a sufficient degree, using state-of-the-art privacy-preserving machine learning techniques.[236] Gatekeepers should have to document their choice of the anonymization technique and the reasons for choosing it.[237] This would ensure that gatekeepers have to strike an explicit, documented, and auditable balance between safeguarding data protection rights of affected persons and the interests of competitors in receiving useful data sets.

### 3. Fairness

Our third proposal relates to rankings. They are now at the heart of the digital economy. Rankings are the logical answer to the digitally mediated, excessive supply of information and products.[238] By their ordering and prioritizing effect, they control demand, search and buying behavior.[239] They represent the core of the business model of many gatekeepers and are therefore rightly regulated even more comprehensively in the final version of the DMA than in the Commission's draft. In particular, the expansion to crawling and indexing seems sensible, because this can decisively influence the visibility of products and thus, for example, the rankings created by search engines (cf. Recital 51 DMA). The explicit inclusion of virtual assistants as possible creators of rankings (Articles 6(5) and 2(22) DMA) is also fully justified.

Already now, classical anti-discrimination law is in principle applicable to rankings.[240] However, it often falls short, since the protected characteristics listed there, such as gender or ethnic origin, generally only refer to persons and not to objects. Therefore, in principle, it is necessary to operate, in the area of e-commerce, with a broadened equal treatment rule. This provides an important building block for fairness in e-commerce.[241] Article 6(5) DMA further develops the case constellations known from competition law. However, the rule must remain operationalizable for gatekeepers, especially in view of the significant threat of sanctions. Furthermore, the economic core function of rankings, to enable the realization of preferences through selective ordering, must not be undermined. At the same time, however, the competitive effects of rankings must be considered precisely because of their selection and steering effects.

---

[235] See, e.g., Joseph Sakshaug et al., 'The effect of framing and placement on linkage consent' (2019) 83 Public Opinion Quarterly 289.
[236] See n. 191 and accompanying text.
[237] Cf. Art. 11 AI Act.
[238] Racula Ursu, 'The Power of Rankings: Quantifying the Effect of Rankings on Online Consumer Search and Purchase Decisions' (2018) 37 Marketing Science 530, 530.
[239] Racula Ursu, 'The Power of Rankings: Quantifying the Effect of Rankings on Online Consumer Search and Purchase Decisions' (2018) 37 Marketing Science 530, 549.
[240] Philipp Hacker, 'Teaching fairness to artificial intelligence: existing and novel strategies against algorithmic discrimination under EU law' (2018) 55 Common Market Law Review 1143, 1159.
[241] On the concept of fairness in the DMA, see Heike Schweitzer, 'The Art to Make Gatekeeper Positions Contestable and the Challenge to Know What Is Fair: A Discussion of the Digital Markets Act Proposal' (2021) 29 Zeitschrift für Europäisches Privatrecht 2021, 503; Rupprecht Podszun, Philipp Bongartz and Sarah Langenstein, 'The Digital Markets Act: Moving from Competition Law to Regulation for Large Gatekeepers' (2021) 11 EuCML 60, 62; see also Wolfgang Fikentscher, Philipp Hacker, Rupprecht Podszun, FairEconomy, (Springer 2013).

This is epitomized by currently hotly debated popularity-based rankings, according to which the order is defined by the presumed attractiveness of the items to users.[242] On the one hand, a differentiation of ranked products according to the expected purchase and click rate does provide a feasible shortcut for approximating rankings to user preferences. On the other hand, the technique may have anti-competitive effects insofar as products that have already been on the market for a longer time tend to be favored over new ones, since only the former can demonstrate a successful historical purchase and click rate.[243] This may entrench the position of historically successful brands and companies to the detriment of newcomers. Therefore, it seems reasonable to compel gatekeepers, by way of a teleological interpretation of Article 6(5) DMA, to shuffle their popularity rankings, for example by reserving some attractive ranking positions for new products.[244] In this way, in our view, the interest of consumers in receiving product recommendations that are as predictive of preferences as possible could be combined with the interest in dynamic competition.

The same can be said for voice commerce, which is essential especially in the area of virtual assistants. Here, usually only one product is selected, which is ordered via voice control.[245] Behind this selection, however, is a ranking,[246] which is typically only visible in the app.[247] This ranking must also conform to Article 6(5) DMA for gatekeepers. Recital 48 DMA now makes this unequivocally clear. However, since a single product is highlighted, the gatekeeper ought to ensure a shuffling of this highlighted position to comply with Article 6(5) DMA. This could be achieved, for example, by permuting (across all comparable queries) the items occupying the top three ranking positions: each of the three highest ranked elements criteria is randomly placed in the top position for one third of the queries. The underlying ranking must, of course, be transparent, fair and non-discriminatory as per Article 6(5) DMA. In this way, catering to consumer preferences is combined with a technologically mediated process for fostering workable competition and preventing winner-takes-all markets.[248]

---

[242] See, e.g., Yotam Shmargad and Samara Klar, 'Sorting the news: How ranking by popularity polarizes our politics' (2020) 37 Political Communication 423; Fabrizio Germano et al., 'The few-get-richer: a surprising consequence of popularity-based rankings?' (2019) The World Wide Web Conference 2764.
[243] Brouwer, 'Privacy self-management and the issue of privacy externalities: of thwarted expectations, and harmful exploitation' (2020) 9 Internet Policy Review 1, 17.
[244] Sandeep Pandey et al., 'Shuffling a Stacked Deck: The Case for Partially Randomized Ranking of Search Engine Results' (2005) Proceedings of the 31st VLDB Conference, DOI:10.48550/arXiv.cs/0503011.
[245] Ariel Ezrachi and Maurice E. Stucke, 'Is your digital assistant devious?' Oxford Legal Studies Research Paper 52/2016.
[246] Alex Mari, 'Voice Commerce: Understanding Shopping-Related Voice Assistants and their Effect on Brands', (IMMAA Annual Conference, 2019) 4.
[247] Christoph Busch, 'Mehr Fairness und Transparenz in der Plattformökonomie? Die neue P2B-Verordnung im Überblick' (2019) 121 Gewerblicher Rechtsschutz und Urheberrecht 788, 792.
[248] Cf. Björn Kuchinke and Miguel Vidal, 'Exclusionary strategies and the rise of winner-takes-it-all markets on the Internet' (2016) 40 Telecommunications Policy 582.

## VII. Summary

The DMA will not only provide more competitive opportunities on, alongside, and between large online platforms, but will also decisively shape the way gatekeepers and their competitors deal with AI. Especially in the digital economy, the DMA's impact on AI systems is likely to be much more noticeable than that of the future AI Act, unless its list of high-risk applications is significantly expanded until its enactment.

The provisions of the DMA relevant for AI can be divided into four areas. First, new rules for fair rankings are introduced. With this, the DMA ventures into the core of the AI-based business model of gatekeepers. Article 6(5) DMA consolidates the prohibition of self-preference known from competition law and transparency rules for rankings already existing in other EU law instruments. However, the inclusion of F(R)AND criteria for rankings is new and potentially groundbreaking. They point significantly beyond existing anti-discrimination law and, in our view, introduce a need for justifying differentiations between comparable products in the ranking. From a technical point of view, techniques developed in the computer science research on algorithmic fairness can be used. However, adapting this framework to the DMA is complex due to the potentially unlimited number of protected attribute combinations–unlike in classical anti-discrimination law. The compliance requirements must take this into account.

Second, the use of data and thus also in particular its collection and use for AI training by gatekeepers are significantly restricted. The thrust here is diametrically opposed to that of Article 10 AI Act. To put it bluntly: the DMA does not, in contrast to the AI Act, seek to foster high-performing AI, but to prevent additional improvements by gatekeepers' models based on the specific competitive setting in which gatekeepers operate. Third, access rights are created for business users to enable them to develop high-performance AI models themselves. Fourth, the DMA harnesses information obligations to reduce the information asymmetry between gatekeepers and their business users, especially in the area of advertising.

We complement these findings with policy suggestions in three main areas. First, the AI Act should spell out a coherent and precise transparency regime. It must clarify the relationship to various technical strategies to implement explainable AI, and take trade secrets and manipulability of rankings into account by a smart design of disclosures. Second, access rights for users need to be expanded and data protection safeguards be balanced with the interests of gatekeepers' competitors, and of society at large, in the provision of meaningful data sets that do allow for the development of products contesting gatekeeper positions. Third, fair rankings also need to balance the rankings' original economic function–selecting items and thus facilitating the fulfillment of consumer preferences–with the broader competitive interest in preventing winner-takes-all markets in which newcomers fight an uphill battle to climb in popularity-based rankings.

All in all, the paper shows that the DMA seeks to bridge a variety of economic and non-economic discourses and combines crucial societal interests that necessitate delicate balancing exercises at many points. The recently enacted regulation furthermore points to currently under-explored questions, at the intersection of law and computer science, surrounding the optimal degree of transparency and fairness of e-commerce rankings–one of the key competition ingredients of the digital economy.